\documentclass[prx, 
reprint,
superscriptaddress,
amsmath,amsfonts,amssymb]{revtex4-2} 
\usepackage{graphicx}
\usepackage{comment}
\usepackage{hyperref}

\usepackage[dvipsnames,usenames]{xcolor}

\begin{document} 

\title{Broadband Optical Modulation and Control at Millikelvin Temperatures}
\author{N. Tabassum}
\affiliation{SLAC National Accelerator Laboratory, Menlo Park, California 94025, USA}
\affiliation{Kavli Institute for Particle Astrophysics and Cosmology, Stanford University, Stanford, CA 94035, USA}

\author{T.~Aralis}
\email[]{taralis@slac.stanford.edu}
\affiliation{SLAC National Accelerator Laboratory, Menlo Park, California 94025, USA}
\affiliation{Kavli Institute for Particle Astrophysics and Cosmology, Stanford University, Stanford, CA 94035, USA}

\author{J.~Anczarski}
\affiliation{SLAC National Accelerator Laboratory, Menlo Park, California 94025, USA}
\affiliation{Kavli Institute for Particle Astrophysics and Cosmology, Stanford University, Stanford, CA 94035, USA}

\author{D. Baxter}
\affiliation{Fermi National Accelerator Laboratory, Center for Particle Astrophysics, Batavia, IL 60510 USA}

\author{B.~Cabrera}
\affiliation{SLAC National Accelerator Laboratory, Menlo Park, California 94025, USA}
\affiliation{Kavli Institute for Particle Astrophysics and Cosmology, Stanford University, Stanford, CA 94035, USA}

\author{R. Chapla}
\affiliation{Santa Clara University, Santa Clara, CA 95053, USA}

\author{N.~Entin}
\affiliation{SLAC National Accelerator Laboratory, Menlo Park, California 94025, USA}
\affiliation{Kavli Institute for Particle Astrophysics and Cosmology, Stanford University, Stanford, CA 94035, USA}

\author{L. Hsu}
\affiliation{Fermi National Accelerator Laboratory, Center for Particle Astrophysics, Batavia, IL 60510 USA}

\author{H.W.~Magoon}
\affiliation{SLAC National Accelerator Laboratory, Menlo Park, California 94025, USA}
\affiliation{Kavli Institute for Particle Astrophysics and Cosmology, Stanford University, Stanford, CA 94035, USA}

\author{A.~Nunez}
\affiliation{SLAC National Accelerator Laboratory, Menlo Park, California 94025, USA}
\affiliation{Kavli Institute for Particle Astrophysics and Cosmology, Stanford University, Stanford, CA 94035, USA}

\author{G.~P\'erez}
\affiliation{SLAC National Accelerator Laboratory, Menlo Park, California 94025, USA}
\affiliation{Kavli Institute for Particle Astrophysics and Cosmology, Stanford University, Stanford, CA 94035, USA}

\author{J.L.~Ryan}
\affiliation{SLAC National Accelerator Laboratory, Menlo Park, California 94025, USA}
\affiliation{Kavli Institute for Particle Astrophysics and Cosmology, Stanford University, Stanford, CA 94035, USA}

\author{M.~Salatino}
\affiliation{Kavli Institute for Particle Astrophysics and Cosmology, Stanford University, Stanford, CA 94035, USA}

\author{A.~Simchony}
\affiliation{SLAC National Accelerator Laboratory, Menlo Park, California 94025, USA}
\affiliation{Kavli Institute for Particle Astrophysics and Cosmology, Stanford University, Stanford, CA 94035, USA}

\author{Z.J.~Smith}
\affiliation{SLAC National Accelerator Laboratory, Menlo Park, California 94025, USA}
\affiliation{Kavli Institute for Particle Astrophysics and Cosmology, Stanford University, Stanford, CA 94035, USA}

\author{S.~Stevens}
\affiliation{SLAC National Accelerator Laboratory, Menlo Park, California 94025, USA}
\affiliation{Kavli Institute for Particle Astrophysics and Cosmology, Stanford University, Stanford, CA 94035, USA}

\author{H.~Stueber}
\affiliation{SLAC National Accelerator Laboratory, Menlo Park, California 94025, USA}
\affiliation{Kavli Institute for Particle Astrophysics and Cosmology, Stanford University, Stanford, CA 94035, USA}

\author{B.A. Young}
\affiliation{Santa Clara University, Santa Clara, CA 95053, USA}

\author{N.A. Kurinsky}
\affiliation{SLAC National Accelerator Laboratory, Menlo Park, California 94025, USA}
\affiliation{Kavli Institute for Particle Astrophysics and Cosmology, Stanford University, Stanford, CA 94035, USA}

\author{K. Stifter}
\email[]{kstifter@slac.stanford.edu}
\affiliation{SLAC National Accelerator Laboratory, Menlo Park, California 94025, USA}
\affiliation{Kavli Institute for Particle Astrophysics and Cosmology, Stanford University, Stanford, CA 94035, USA}
\date{\today}

\begin{abstract}
A universal experimental challenge in studying radiation effects on cryogenic devices is precisely and accurately characterizing the position-dependent device response near the energy detection threshold. We have developed a compact cryogenic optical beam steering system that can be used to generate O($\mu$s) pulses of small numbers of photons over the energy range of 1.3 - 3.4\,eV at room temperature, and deliver those photons via fiber optic to any specified location on the surface of a detector operating at cryogenic temperatures. This new system will allow for robust calibration of any photon-sensitive detector, including superconducting devices. The system can be used efficiently to explore the physics of target materials, quantify the position sensitivity of different sensor designs, measure phonon transport, and study the effects of quasiparticle poisoning on detector operation. We describe the design of this pulsed calibration method and present first results obtained with a second-generation system operated at room temperature and sub-Kelvin temperatures. 



\end{abstract}

\maketitle

\section{Introduction}

The radiation response of cryogenic devices is an experimentally interesting question for a number of fields. For telescope applications, it is usually suitable to test sensors in a built environment with a fake focal plane. For the newest generation of far-IR telescopes, however, room-temperature radiation precludes the use of sources above 4~K, and 4~K backgrounds represent a critical background themselves (see e.g. Refs~\cite{echternach2018,day202425micronsinglephotonsensitive}). For other sensitive applications, such as operating high-coherence qubits or rare-event detectors, calibrating spatial response of devices comes into conflict with the requirement to optically isolate these devices from the outside world. External radiation sources can be used to provide controlled radiative doses, but these events are often both too energetic and randomly positioned to help inform detailed models of charge and energy transport (see e.g. Refs~\cite{larson2025quasiparticlepoisoningsuperconductingqubits,hull2025characterizationtesbasedanticoincidencedetector}). Optical fibers can be used to deliver more precisely targeted photons, but without additional components to move the optical beam, only limited spatial information can be gleaned from a given experimental setup (see e.g. Refs\cite{Hong_2020,anthonypetersen2024lowenergybackgroundsexcess,wolfe2024controlthresholdvoltagessisige}). Some limited control over X-ray and optical sources can be achieved with commerically available peizo-electric stages, but these lead to substantial heating during each movement (see e.g. Ref~\cite{huang2024graphenecalorimetricsinglephotondetector}), limiting scan fidelity and consistency between measurements.

Some of the authors of this paper have historically been involved in developing this MEMs mirror-based technique for cryogenic optical control, first at 0.5~K\cite{Moffatt_2016,Moffatt:2016kok,Moffatt_2019,Stanford_2020,Stanford_2021} in a 4~K photon bath, and more recently isothermally at 10~mK\cite{stifter2024}. All of these systems notably used custom-built refractive fiber focusers, with the majority of measurements taking place at a single wavelength, and used monochromatic laser or LED sources. For the final measurement with the 0.5~K system, the photo-electric cross-section of Si was measured as a function of wavelength, which required swapping in a new focuser for each measurement\cite{Stanford_2021}. This data was nonetheless very useful, and led to a desire for an intrinsically broad-band system that could provide monochromatic operation over a wide wavelength range by simply swapping out LEDs at room temperature.

In discussing the need for this broad-band system, the general architecture of a multiple-mirror system was developed, in which a first mirror could generate temporal pulses from a filtered broad-band source, and a second mirror could steer those pulses to create a spatially and temporally resolved map of device response. 

\begin{figure*}[!ht]
    \centering
\includegraphics[width=0.45\textwidth]{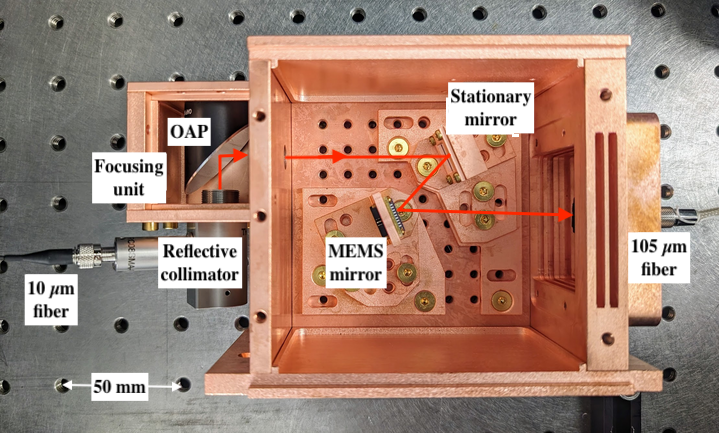}
\includegraphics[width=0.53\textwidth]
    {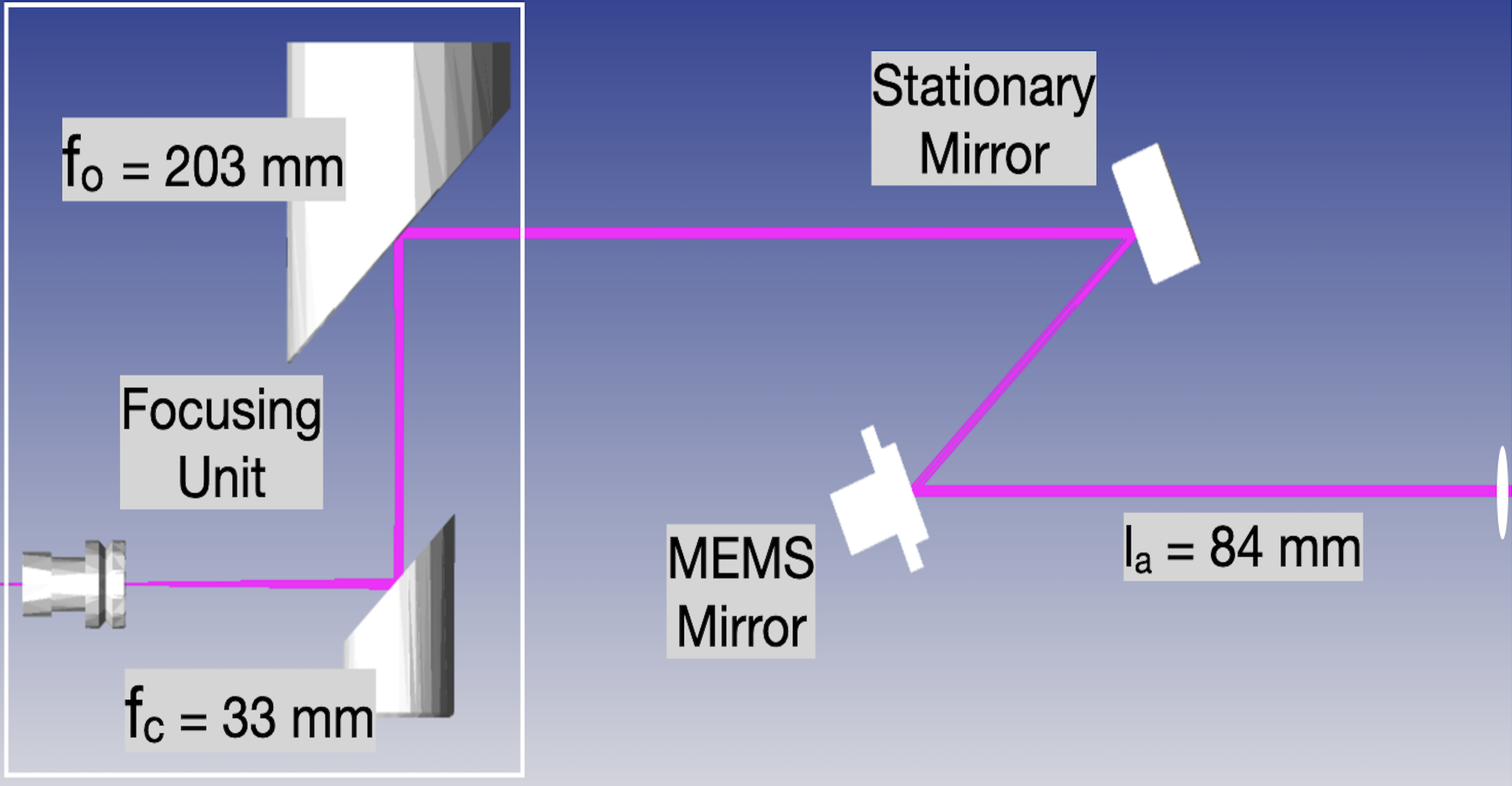}
    \caption{{\bf Left}: Optical chopping test setup (without its copper enclosure ``lid"), shown on the test bench. The reflective focusing unit brings monochromatic photons entering the package via fiber optic  onto a stationary mirror that directs the beam onto a tilt-able MEMS mirror of diameter $d_r$ =~4.6 mm. The voltage-controlled MEMS mirror is then used to scan the photon beam over an exit slit or fiber with a diameter of 105 um mounted to a removable ``output plate". The result is a focused, chopped beam of light on a device-under-test (not shown). {\bf Right}: Simplified diagram (not to scale) of the optical path inside the MEMS scanner unit. The focusing unit includes a reflective collimator unit of focal length $f_c$ = 33 mm and a separate off-axis parabolic mirror (OAP) of focal length $f_0$ = 203 mm. The center of the MEMS mirror is located a distance $l_a$ = 84 mm away from the exit port.}
    \label{fig:diagram}
\end{figure*}

In this paper, we document the first development steps towards this concept, in which we have developed a broad-band scanner/optical chopper and measured its performance across about a decade in wavelength, from 280~nm to 970~nm. In Section~\ref{sec:design}, we describe the design modifications required to move from refractive optics to broad-band reflective optics, based around a small multi-mode input optical fiber. We summarize the expected optical performance when using this device to couple light back into a larger multi-mode output fiber. In Section~\ref{sec:beamSize}, we report on the optical performance of the focusing system, both on its own as well as when installed in the scanner and re-coupled into an output fiber, the latter case including tests at 300~K and 10~mK. Finally, we present 300~K and 10~mK measurements of the system used as an optical chopper, demonstrating a wavelength-independent pulse width below 5~$\mu$s. We conclude by discussing next steps for this system, including extension to longer wavelengths for future implementations and continued reduction in focal point spot size. Appendices A-D contain additional details about the scanner design and performance metrics, as well as details of the basic models used to compute quantities reported in the main text.

\section{MEMS Scanner Design}\label{sec:design}
The cryogenic, MEMS-based photon scanning system used for this work is shown in Figure~\ref{fig:diagram}. The main unit contains three key components: a reflective or refractive focusing unit assembly, a stationary mirror, and a tilt-able MEMS mirror. The system is coupled to a room-temperature photon source via fiber optics. To ensure adequate initial alignment to the external laser system, the MEMS unit and the stationary mirror are separately attached to dedicated precision mounts that provide each component with three translational plus one rotational degrees of freedom. An attractive feature of MEMS mirrors is that they work well at all temperatures between our 10~mK design temperature and 300~K.

The specific system shown in Figure~\ref{fig:diagram} and described in this paper is similar to the previous version presented in Ref~\cite{stifter2024}, but the new unit includes several critical upgrades and uses a reflective collimator unit in place of the original refractive optics.  The design modifications that led to the new system were driven by two main goals:
(i) to provide broad-band performance and remove the need for wavelength-specific focusing optics, and (ii) to enable careful exploration of the system at $\sim 20~mK$ when using chopped and/or swept beams over a range of optical and near-IR wavelengths. 

These requirements, combined with the desire to have a relatively long lever arm to maximize scan range across a detector mounted at the exit port of the full unit, preclude the use of refractive microscope objectives. In addition, the requirement for the input to couple to a cryogenic-friendly, SMA-terminated fiber drove us away from integrated magnifier designs used for microscopy, towards a separate collimator and focusing system that can provide greater design freedom. In the final scanner design shown in Figure~\ref{fig:diagram}, the focusing unit includes a reflective fiber collimator mounted $\sim$ cm away from a parabolic mirror. The role of the stationary mirror shown in the figure is to extend the optical path-length within the compact unit, and redirect the focused beam onto the MEMS mirror.

Below, we describe the recent optimization and enhanced modularity of the focusing unit. We show that the new system produces smaller-diameter and more uniform photon-beam spots than our original design, while maintaining the excellent scan-range characteristics of the older system.  We then describe recent design modifications to the final output stage of the scanner system where it couples to the device under test, and discuss the significance of these modifications.

\subsection{New Broadband Focuser}


The new MEMS scanner focusing unit was designed to replace a refractive-based focuser unit with a drop-in, reflective unit that would not require re-design of the full MEMS platform. Taking this approach ensured that our existing systems could be easily retrofitted for broad-band operation. As shown in Appendix~\ref{app:timing}, when the MEMS mirror is ignored, the focusing unit itself is just a magnifier, and the beam spot size is minimized by maximizing the effective focal length of the collimator for a fixed output focal length. In addition, when this system is used as an optical chopper, the output focal length effectively cancels out of the pulse width equation, and
the pulse duration essentially becomes determined by the collimator focal length alone.

To accommodate the existing 
MEMs scanner footprint, the new reflective-optics design uses a Thorlabs MPD189-P01, 203.2~mm focal length off-axis parabolic mirror (OAP) as the primary focusing optic. A Thorlabs RC08 reflective collimator of focal length $f_c$ = 33~mm and maximum numerical aperture $NA$ of 0.15 is used to focus the fiber output onto the OAP. The overall design yields a beam magnification factor $M$ of: 
$$
M=\frac{d_{out}}{d_{in}}=\frac{f_0}{f_c}=\frac{203.2}{33}\approx 6.2.
$$

As configured, the new focusing unit provides a maximum collimated beam size of 11~mm for input fibers 0.15 NA or larger, which is approximately twice the diameter ($d_r$ = 4.6 mm) of our MEMS mirror (Mirrocle{\textsuperscript{TM} model A8L2.2).
As a result, the maximum lever arm for the new scanner design was set to approximately one-half the focal length of the OAP focuser. 

Using equation~\ref{eq:spotSize} in appendix~\ref{app:timing}, we show that the effective beam spot size $d_{eff}$ at the output port of the scanner is expected to scale as:
\begin{equation}
    d_{eff} \approx 6.2\sqrt{d_{f}^2+(1.22\lambda/(2NA))^2}.
\end{equation}
Thus, a multi-mode input fiber of diameter $d_f=10~\mathrm{\mu m}$ and $NA\sim 0.1$ should produce a divergence-limited spot diameter $d_{eff}\approx 62~\mathrm{\mu m}$.
If one includes diffraction effects, the expected spot size increases to:
\begin{equation}\label{eq:spotSizeEstimate}
    d_{eff}' \approx 62~ \mathrm{\mu m}\sqrt{1+(0.61\lambda\cdot 10^6)^2}
\end{equation}
This result satisfies our initial goal of adopting the path length  geometry of our existing system to achieve $100~\mathrm{\mu m}$ or smaller spot diameters in the optical regime at the location of the exit port. In comparison, the refractive system of Ref~\cite{stifter2024} achieved a spot size 170~$\mu$m, so we expect this new system to improve on that by a factor of just under 3.

A second goal was to provide a scan range $d_{scan}$ $>$~20~mm. As shown in Equation~\ref{eq:scanRange_app} of Appendix~\ref{app:timing}, the new collimation setup gives: 
\begin{equation}
    d_{scan} \lesssim 49~\mathrm{mm} \left[\frac{d_r}{4.6~\mathrm{mm}}\right] \left[\frac{\theta}{\theta_{max}}\right],
\end{equation}
where $\theta$ is the beam deflection angle off the MEMS mirror ($\theta_{max} \approx \pm5^{o}$) and $d_r$ is the MEMS mirror diameter (here, $d_r$=4.6 mm).

Both the scan range and spot size scale with the magnification factor, for fixed fiber and MEMS mirror specifications. The chosen mirror offers an excellent combination of size and range of deflection angle that, when coupled with the breadboard-based design of the scanner unit, yields small spot sizes for relatively large (and adjustable) scan ranges. 
Details about the trade-offs between scan angle and spot size are described in more detail in Appendix~\ref{app:timing}.



\subsection{Output Fiber Plate}\label{sec:OutputFiberPlate}
In contrast to Ref~\cite{stifter2024}, the characterization data reported in this paper were obtained without a cryogenic detector payload, to best evaluate the MEMS scanner system alone. 
We considered a number of measurement approaches for the characterization work, including, e.g., sweeping the beam across a pinhole much smaller than the beam diameter, etc. However, such approaches would require additional optics and add complexity and alignment uncertainty during cryogenic cycling. In the end, we simply covered the output port with a precision plate that placed a 105~$\mu$m multi-mode fiber at the nominal center position of the scan field (see Figure~\ref{fig:diagram}). The diameter of the output fiber was selected to be as small as possible, while exceeding the expected cryogenic spot size of 60~$\mu$m.


In our model, the output beam shape is treated as a convolution of the beam and output fiber profiles. Assuming a roughly Gaussian beam profile, the effective spot size of the focused beam at the scanner output plate should be:
\begin{align}
    d_{eff,meas} &= \sqrt{d_{eff}'^2+d_{fiber}^2} \\
    &\approx 122~\mathrm{\mu m}\sqrt{1 + (0.31\lambda\cdot 10^6)^2}
\end{align}
which is roughly 130~$\mu$m for a wavelength of $\lambda$ = 625~nm. This value for the spot size is roughly double what we measure with a detector when looking at output light intensity vs. spot position, and would imply $\sim$ twice the relevant sweep length value for pulsed beam calculations, as discussed in Section~\ref{sec:pulseWidth}. 

\begin{figure*}[th]
    \centering
    \includegraphics[width=0.9\textwidth]{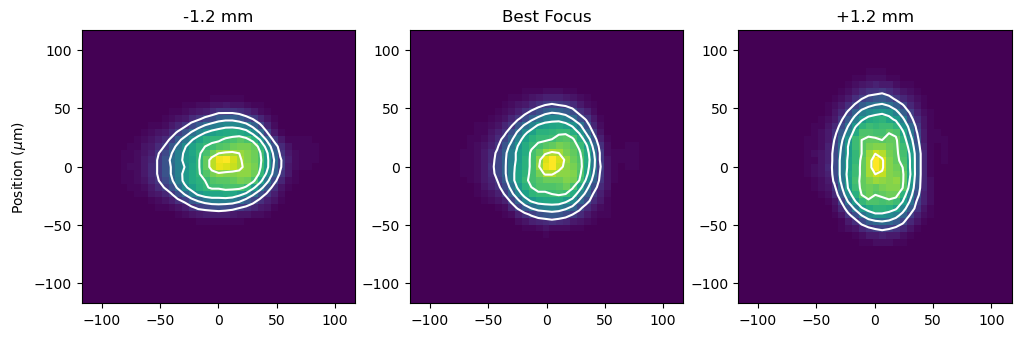}
    \includegraphics[width=0.45\linewidth]{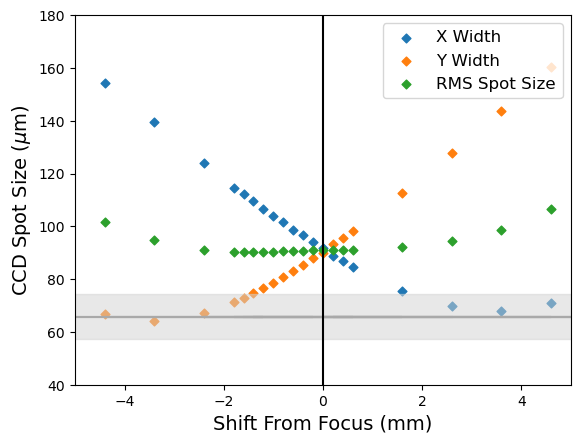}
    \includegraphics[width=0.45\linewidth]{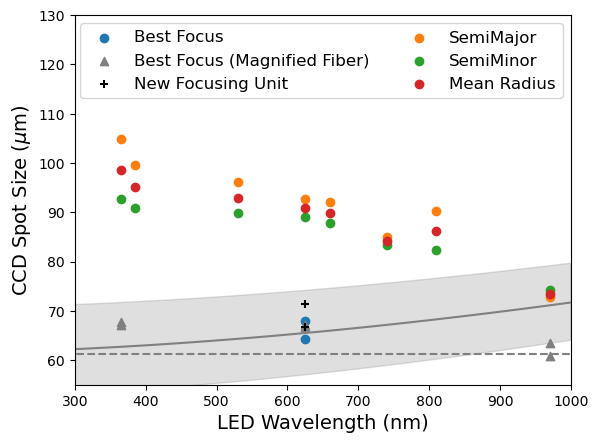}
    \caption{\textbf{Top Row}: Example images of the focused beam spot taken with a 4.8 $\mu$m-pixel CCD camera and $\lambda$ = 625 nm photons. The setup included only the reflective focusing unit (OAP and collimator) and not the MEMS mirror. The central image shows the best focus, where the spot is round with a (4$\sigma$) full width of 90~$\mu$m. The left and right images show the effect on spot size and shape of defocusing the beam by $\pm$1.2 mm, illustrating the slight astigmatism in our system. The source of this astigmatism was later found to be caused by slight angular misalignments within the focusing unit. See Appendix~\ref{app:newFocuser} for details on how this issue was mitigated. \textbf{Bottom Left}: Measurements of the spot size from the reflective focusing unit (only) at 625~nm as a function of distance from the expected focal length, based on the calculated spot size for $M=6.16$, NA=0.1, and mean fiber diameter of 10 $\mu$m. The gray bar is the range of spot sizes expected within the quoted range of Thorlabs fiber diameter, which is only accurate to the 3~$\mu$m level. Astigmatism in the focusing unit is evident in the top plots, where the best focus is located where X and Y have the same width, yet each (X and Y) has a focus shifted relative to this nominal focal point. For a system with astigmatism corrected, we thus expect this system to be diffraction limited at 625~nm to 65~$\mu$m, which is in agreement with the lower range estimate of the theory band. We have demonstrated this correction to the astigmatism and present initial results in Appendix~\ref{app:newFocuser}. \textbf{Bottom Right}: Spot size vs wavelength for the range of optical diodes used in this study, where the mean radius, semi-major and semi-minor axes are as-measured values, obtained in the full MEMS system during warm checkouts. The absolute minimum spot size, set by the magnified size of the 10 micron fiber, is shown as a horizontal dashed line. The best focus shows the spot size at the respective X and Y focuses from the left figure, which are much closer to the divergence limit. The gray points show the imputed spot size using a system with a larger magnification, which reduces diffraction and astigmatism effects. Finally, the black points show the major and minor axes of the spot produced by a newer collimator design that largely corrects the astigmatism seen in this system. The grey band shows the expected diffraction-limited performance assuming a 4$\sigma$ fiber diameter of 10$\mu$m. The fact that the imputed spot size at $1\mu$m is below this band reflects the fact that the fiber is approaching the single-mode limit.}
    \label{fig:spotSizeMeasured}
\end{figure*}

\section{Beam Size Measurements}\label{sec:beamSize}

We conducted both warm (300 K) and cold ($\sim$10 mK) measurements of the beam spot size to validate the design expectations summarized above. 
By design, the scanner performance is relatively insensitive to thermal contraction effects, so we were able to perform most of the system characterization and optimization experiments at room temperature, rather than having to cool the unit after each small refinement was made to the design. Below, we document two types of room temperature beam size tests, where the scanner was either: (i) mounted on an optical bench, and the beam size measured using a CCD detector, or (ii) mounted $in situ$ in the (warm) cryostat and the beam size measured using an avalanche photodiode detector (APD). Unless otherwise noted, all bench-top tests were performed using the same models of input and output fibers as those used for the $in situ$ tests. 

\subsection{CCD Measurements}

Bench-top measurements of beam spot diameter produced by the two-part reflective focusing unit were made using 625-nm photons and a CCD camera with 4.8 $\mu$m pixel pitch (Thorlabs Model CS235MU). The focal plane of the camera was moved through the expected focal point of the full MEMS scanner, using a micrometer stage. Typical results of the CCD measurements are shown as intensity contour plots in the three upper panels of Figure~\ref{fig:spotSizeMeasured}. The ``best-focus" data is shown in the center. Images obtained with the camera $\pm$ 1.2 mm away from the focus along the optical axis are also shown (to the left and right).
The beam spots were characterized by fitting ellipsoidal beam profiles to the observed images, with 4$\sigma$ spot sizes reported in this paper. The beam spot size results are presented in Figure~\ref{fig:spotSizeMeasured} (lower left) and are compared to expected spot sizes computed with Equation~\ref{eq:spotSizeEstimate}. 

We expected that the beam spot size would be limited by our fiber diameter. Instead, we found that our spot sizes were systematically 30-50\% larger than the design values (see the lower right panel of Figure~\ref{fig:spotSizeMeasured}). Figure~\ref{fig:spotSizeMeasured} illustrates that, due to a slight angular misalignment of optics, this focuser has a pronounced astigmatism, with the optimal focus in the tangential and saggital planes located at different focal lengths. By measuring the two-dimensional spot shape as a function of position relative to the nominal focus, we were able to confirm that the minimum beam spot width at the true focus in each plane matched the design value. In Appendix~\ref{app:newFocuser}, we discuss results with a newer focuser design in which this astigmatism is substantially reduced thanks to a unibody design. We found that most of the focusers of this design had a similar astigmatism, suggesting a design defect leading to a consistent misalignment of the OAP and collimator by a few tenths of a degree.

The beam size was also measured as a function of photon wavelength, using a set of nine fiber-coupled LED diodes spanning 365 nm - 970 nm. In these experiments, the camera was placed at the optimal focal point found for 625~nm photons. The results are shown in Figure~\ref{fig:spotSizeMeasured} (bottom right) and Table~\ref{tab:pulseWidth}. This trend is a reflection of the departure from Gaussian optics in our system. When we tested the new focusing unit described in Appendix~\ref{app:newFocuser}, we found that the mode density in the fiber increases rapidly with decreasing wavelength, with $M^2$ approaching 1 at 970~nm. As a result, the effect of astigmatism produces a more pronounced increase in spot size for shorter wavelengths. When only considering focus in one plane, we can see that a similar reduction in spot size is observed at longer wavelength, but the effect is much smaller, in line with the change being driven by simply a reduced mode field diameter as the fiber modes decrease. The spot size of the new focuser at nominal focus lines up well with the optimal focus measurements in the collimator used for this cryogenic test; both agree with the modeled expectation based on a fixed-mode field diameter and uncertainties in fiber properties. Future work should confirm that for single-mode waveguides, we see the expected increase in spot size with wavelength above 1~$\mu$m.




\subsection{APD Measurements}

The beam spot size was also characterized using the output fiber plate and an avalanche photodiode detector (APD). For these measurements, the scanner was mounted inside the cryostat and coupled to {\it in situ} input and output fibers, as discucsed in detail in Appendix~\ref{app:expSetup}. With the MEMS mirror in stationary mode, 625 nm photons were sent through the system, and the relative intensity of the output beam was measured using the mean value of the measured APD signal (voltage). The 2-axis MEMS mirror was then tilted using DC voltage control to a new (stationary) orientation, and the averaged APD signal was recorded again. The process was repeated at a number of mirror positions to obtain a full image of the beam spot intensity over two dimensions.

\begin{figure}[!t]
    \centering
    \includegraphics[width=0.45\textwidth]{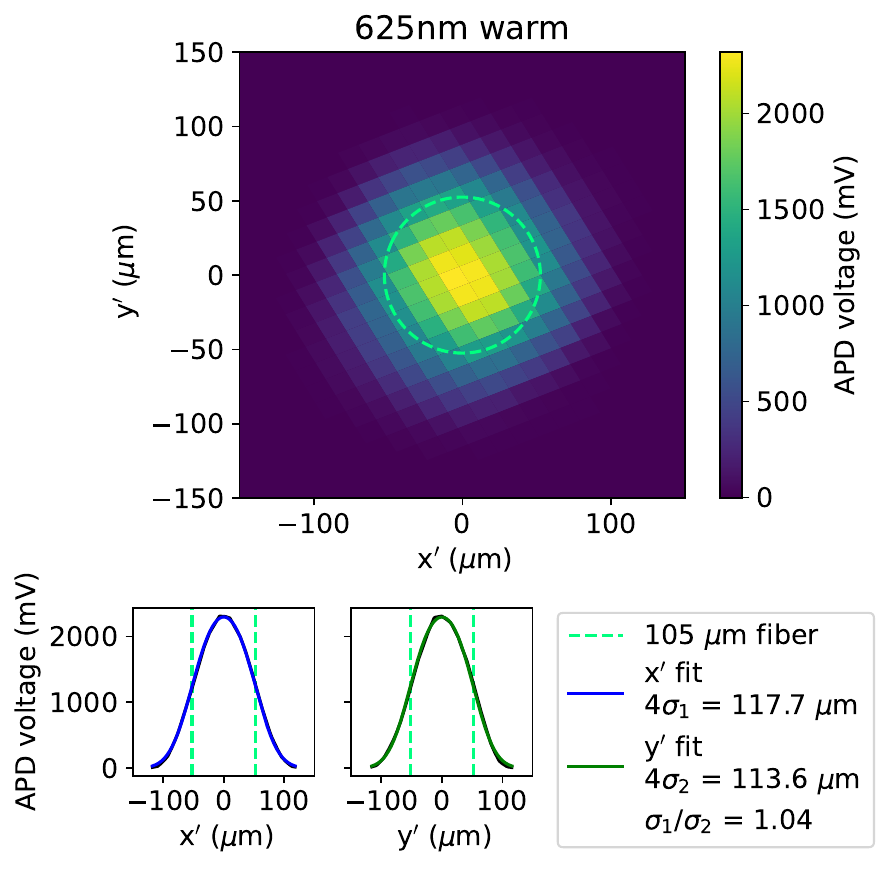}
    \caption{\textbf{Top}: Room temperature image of beam coupled to 105~$\mu$m diameter output fiber. Intensity through the output fiber was measured using voltage out of an APD. The conversion from 2-axis MEMS tilt controller settings to physical units ($\mu$m) assumes the intensity profile along each principle component of the spot is equivalent to a Gaussian convolved with a 105~$\mu$m square pulse. \textbf{Bottom Left}: The intensity along one principal component fit to a Gaussian convolved with a 105~$\mu$m square pulse. \textbf{Bottom Center}: Same as Bottom Left for the second principle component. \textbf{Bottom Right}: The relevant legend, 
    including 4$\sigma$ widths of the relevant Gaussian fits.}
    \label{fig:APDspotSizeWarm}
\end{figure}

Images were then converted from dimensionless MEMS tilt-control units to distances using the known diameter of the output fiber (105~$\mu$m) as a calibration feature, and assuming each measured image was a convolution of the circular fiber geometry with a bivariate Gaussian beam profile (see Appendix~\ref{app:spotCalibration}). An example beam spot image taken on the bench with this method can be seen in Figure~\ref{fig:APDspotSizeWarm}. In this example, the predicted fiber-removed spot size is $\sim$30\% larger than that observed with the CCD. We attribute the difference to larger systematic uncertainties associated with the MEMS-tilt method. Despite these uncertainties, the tilt-based measurement scheme is useful because it enables {\it in situ} beam spot measurements over a broad temperature range.

\begin{figure*}[t]
    \centering
    \includegraphics[width=0.45\linewidth]{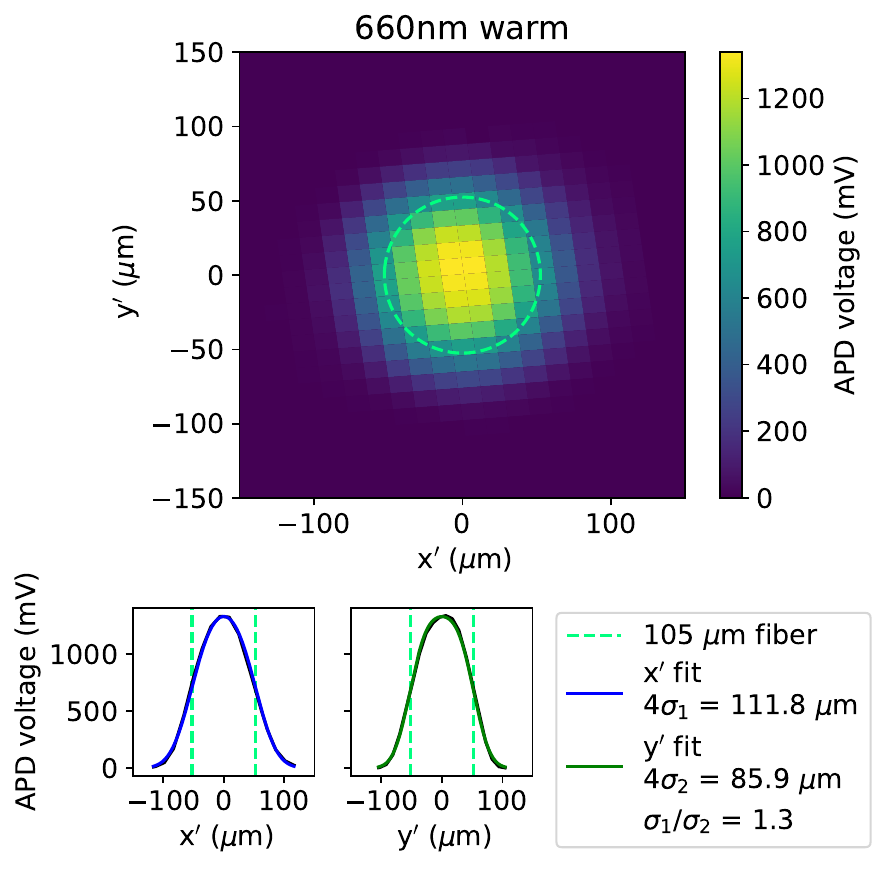}
    \includegraphics[width=0.45\linewidth]{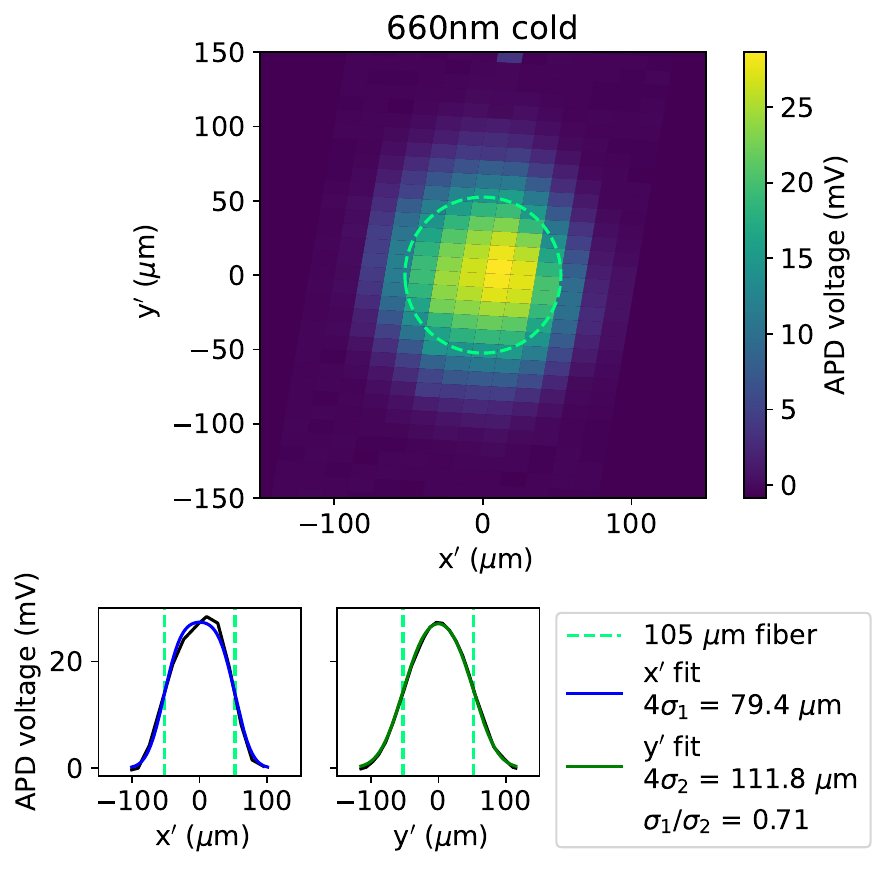}
    \caption{\textbf{Left}: Room-temperature image of 660~nm beam coupled to 105~$\mu$m diameter output fiber. Individual figures comparable to those in Figure~\ref{fig:APDspotSizeWarm}. \textbf{Right}: Cryogenic (10 mK) 660~nm image.}
    \label{fig:APDspotSize660}
\end{figure*}

It was found that spot size results did not change appreciably between warm and cold scanning measurements, so the majority of {\it in situ} characterization data were obtained at 660~nm, as the LED source at this wavelength provided the best signal-to-noise ratio when run through the full set of fibers in the cryostat.  Figure ~\ref{fig:APDspotSize660} shows for comparison 660-nm beam spot images obtained at 300~K and 10~mK, using the fiber-plate and APD readout scheme. Cross-checks were performed at 365~nm to test for any wavelength dependence; the differences found were even smaller than those observed in bench-top tests performed with the focuser alone. 

As shown in Figure ~\ref{fig:APDspotSize660}, when the scanner was cooled to 10 mK, the RMS spot-size decreased by $\sim$3\%, and the ratio of spot diameters along perpendicular axes changed by $\sim$50\%. The latter effect is consistent with a reduction in the total optical path-length of $\sim$2~mm and the measured astigmatism shown in Figure~\ref{fig:spotSizeMeasured}. A $\sim$2~mm thermal contraction is within expectations for an initial 203~mm beam length constrained by an OFHC copper housing cooled to 10~mK.

\section{Pulse Width Measurements}\label{sec:pulseWidth}
To produce optical pulses with this device to characterize cryogenic MEMS performance, we drove the MEMS mirror along one axis to sweep the focused beam spot across the center of the optical fiber mounted to the exit plate. We tested a variety of drive waveforms to study the pulse width and shape in each case. While all cases resulted in a roughly Gaussian beam shape, we found that driving the mirror with a square wave produced the shortest pulse duration, consistent with the highest angular beam speed. A low-pass filter included in the MEMS driver prevented oscillation of the MEMS mirror at its resonant frequency, which could damage the mirror. Additional low-pass filtering was provided implicitly by $\sim$ 2~m of twisted pair wiring used to connect the driver to the cryogenic system. The net result of this driving technique contributes to the uncertainty in MEMS angular frequency produced by the drive signal. Determining the actual frequency was one goal of our cryogenic chopping measurements.

For comparison, we can compute our expectation for a pulse generated by a harmonically deflected MEMS mirror along one axis at a rate $\omega$.  As shown in Appendix~\ref{app:timing}, for this geometry, the effective pulse time $t_{pulse}$ for transmitted pulses can be written as:
\begin{equation}\label{eq:pulseWidthRealModel}
    t_{pulse} \approx \frac{d_{eff,meas}}{v_{beam}} = 122\frac{\sqrt{1 + (0.31\lambda\cdot 10^6)^2}}{2\theta_m \omega l_a}
\end{equation}
where $\theta_m$ is the maximum MEMs mirror deflection angle, and $l_a$ is the distance between the MEMS mirror and the output spot. (See Figure~\ref{fig:diagram}.)
The pulse length can be minimized by making $l_a$ as large as possible, while keeping the converging beam spot within the MEMS mirror diameter $d_r$, giving:
\begin{equation}
    t_{pulse} \geq \frac{122 \mathrm{\mu m}}{2\theta_m \omega d_r}\sqrt{1 + (0.31\lambda\cdot 10^6)^2}\frac{2NA}{M}
\end{equation}
All of the parameter values in this equation are known with one exception; we have only a notional idea of the maximum safe angular drive frequency $\omega$ of the device when cold. As a proxy, we use the 300~K value of the MEMS resonant frequency provided by the manufacturer: $\omega_o/2\pi \sim$~330 Hz\footnote{The specification sheet for this mirror/driver combination can be found at \url{https://www.mirrorcletech.com/pdf/DS/MirrorcleTech_Datasheet_A8L2.2-4600.pdf}}. Of course, driving the MEMS mirror too close to resonance risks damaging or even destroying the unit, so as a precaution we generally operate below $\sim$ 50$\%$ $\omega_o$ by applying a digital 120~Hz low-pass filter as recommended by the manufacturer.


\begin{figure*}[t]
    \centering
    \includegraphics[width=0.45\linewidth]{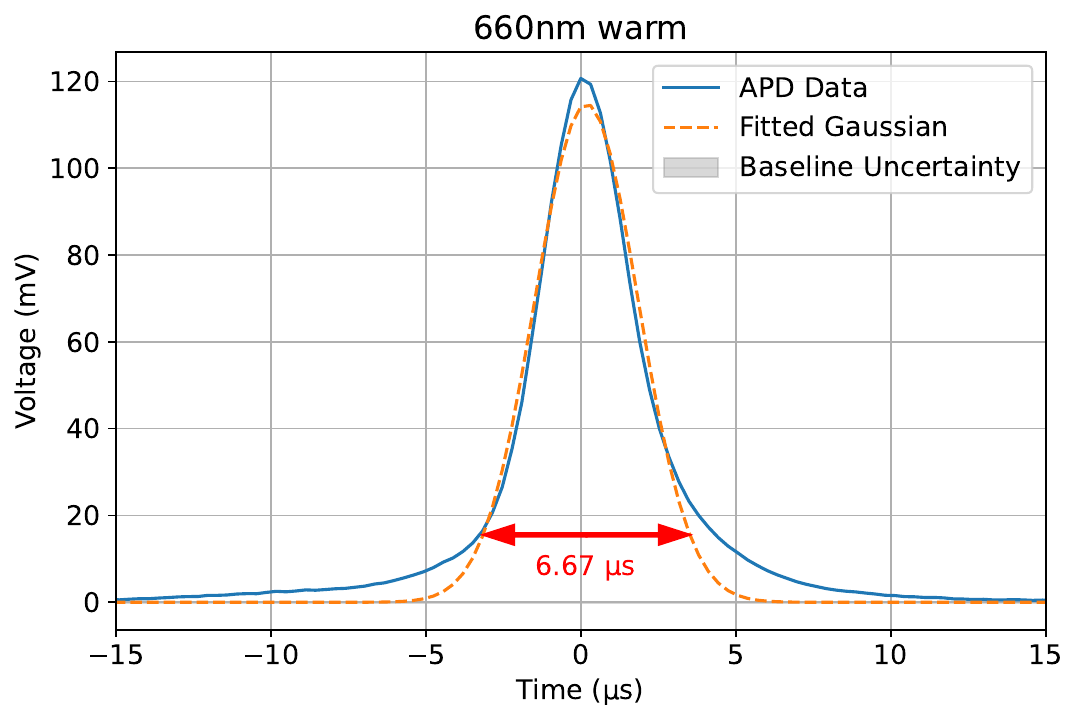}
    \includegraphics[width=0.45\linewidth]{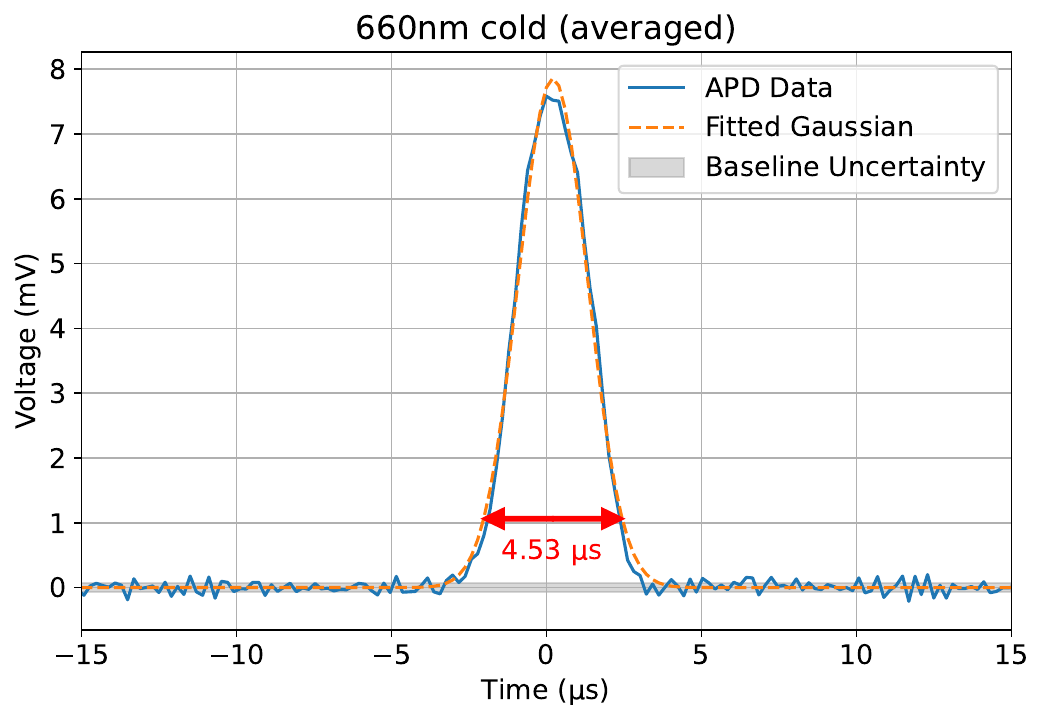}
    \caption{
    \textbf{Left}: Pulse-width measurement from the MEMS chopping system at warm temperature with a 660~nm LED. \textbf{Right}: Pulse width from the MEMS chopping system at cold temperature.}
    \label{fig:pulsesize}
\end{figure*}

Combining the known geometry, component values, and resonant frequency constraint, an estimate of the minimum achievable pulse duration can be readily calculated in terms of LED wavelength $\lambda$ (in nm) assuming oscillation at the resonant frequency, assuming the system is geometrically optimal for pulsed measurement:
\begin{equation}\label{eq:pulseWidthIdeal}
    t_{pulse,ideal} \gtrsim 2.4~\mathrm{\mu s} \sqrt{1 + (0.31\lambda\cdot 10^6)^2}
\end{equation}
The diffraction correction for optical wavelengths is at the percent level, and thus equation~\ref{eq:pulseWidth} implies a minimum pulse width of $\sim$~2.4 $\mu$s at 650~nm for a perfectly optimized drive signal and exactly aligned optics. 

An important caveat is that the mirror locations in the system as tested were defined for larger NA fiber optics used in an earlier version of the scanner. As a result, the distance between the MEMS mirror and the exit plate, $l_a\sim$ 84~mm, is substantially shorter than it could be for optimal performance. Using equation~\ref{eq:pulseWidthRealModel}, we find that for the geometry implemented here, we have
\begin{equation}\label{eq:pulseWidth}
    t_{pulse} \gtrsim 4.0~\mathrm{\mu s} \sqrt{1 + (0.31\lambda\cdot 10^6)^2}
\end{equation}
Thus, even without re-optimization, our existing scanner should be able to produce $\sim4~\mu$s-scale pulse widths. 

Note that equation~\ref{eq:pulseWidth} sets the minimum obtainable pulse width for the current set-up, but pulses can be tuned above this value to arbitrarily slow pulse widths and long duty cycles by simply adjusting the harmonic drive frequency of the MEMS mirror. For the case of the square-wave drive, we are thus testing the highest achievable mirror drive frequency while retaining the filtering needed to produce stable MEMs control. While we expect we could achieve results close to this limiting value by removing hardware filtering, we would risk mirror damage, and opted to keep these limits in place for this test.

To measure the pulse chopping time using the output plate, the y-position of the mirror was set to the centroid of the beam spot and the x-position was set to oscillate with time, driven by the filtered square wave as discussed above. The APD output signals were used as a scope trigger. After signal averaging, the data were fit to a Gaussian profile to extract the 4$\sigma$ pulse width for each wavelength. Example pulse-width measurements made at both 300~K and 10~mK for 660~nm photons are shown in Figure~\ref{fig:pulsesize}. Results from similar measurements made at 10~mK for different $\lambda$ are summarized in Table~\ref{tab:pulseWidth}. We note that the warm measurements are 50--100\% higher than our predicted minimum value of 4~$\mu$s, while the cold values are less than 15\% higher than our predicted limit, with a wavelength-dependent trend consistent with an increase due to the astigmatism effects discussed earlier. We explore the reason for this difference in the next section, as it requires comparing the pulse width to measured spot size to separate focusing and temporal effects.

Due to limitations in performance of the APD and fibers in the IR regime, as well as the measured (in situ) $\sim 7x$ reduction in fiber optic light transmission with the scanner operating at cryogenic temperatures compared to 300~K, cold pulses were indistinguishable from noise for 940-nm and 970~nm sources. 
We note, however, that although the IR pulses were not detected by the APD in our tests, a cryogenic detector coupled to the scanner unit for detector characterization would be sensitive to these smaller energy deposits. We discuss our extensive fiber optic testing performed to cross-check these results in Appendix~\ref{app:expSetup}. We show that the reduction is not due to any property of the reflective optics, but rather to a reduction in transmission in the Thorlabs fibers, due to jacket compression when cold. The $100~\mu$m fiber shows particularly poor temperature dependence and is the leading cause of increased attenuation when cold. We anticipate that by just replacing the Thorlabs fibers with our standard Accuglass fibers would yield a 20x higher SNR than this measurement based on the tests discussed in the appendix.


\begin{table}[!t]
    \centering
      \caption{Spot size and pulse width measurements for MEMS scanner version 2, using CCD and APD (fiber) configurations.} 
    \begin{tabular}{ | c | c | c | c | c | c|}
    \hline
    & \multicolumn{3}{c}{Spot Size [$\mu$m]} & \multicolumn{2}{|c|}{} \\
    & CCD & \multicolumn{2}{c}{Fiber} & \multicolumn{2}{|c|}{Pulse Width [$\mu$s]} \\
    Wavelength (nm) & 300K & 300K & 10 mK & 300K & 10 mK\\ 
    \hline
    365 & 98.6 & & & 7.86 & 4.58 \\
    385 & 95.1 & & & 8.36 & 4.45 \\
    530 & 92.9 & & & 7.01 & 4.39 \\
    625 & 90.9 & 115.7 & & 7.71 & 4.29 \\
    660 & 89.8 & 99.7 & 96.9 & 6.67 & 4.53 \\
    740 & 84.2 & & & 7.16 & 4.41 \\
    810 & 86.2 & & & 5.23 & 4.27 \\
    940 & - & & & 5.41 & - \\
    970 & 73.5 & & & 6.14 & - \\
    \hline
    \end{tabular}
    \label{tab:pulseWidth}
\end{table}

\section{Angular Scan Speed}

In the previous sections, we explored the effect that departures from Gaussian optics had on scanner optical performance, and presented pulse width measurements. Here, we explore the only notable temperature-dependent effect that results in a marked reduction in pulse duration between warm and cold tests. The reduction in pulse width as a function of wavelength in the 10~mK data is entirely explained by the change in spot size observed with the CCD, which means that the pulse width change is not an optical effect. Rather, it is due to a marked change in MEMS mirror responsivity. 

\begin{figure*}[t]
    \centering
    \includegraphics[width=0.45\linewidth]{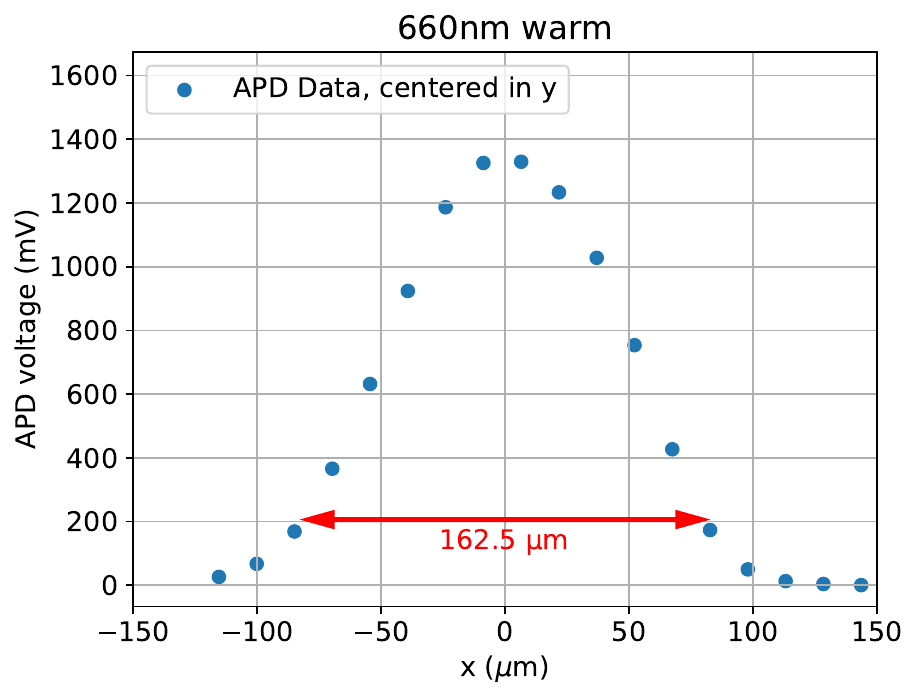}
    \includegraphics[width=0.45\linewidth]{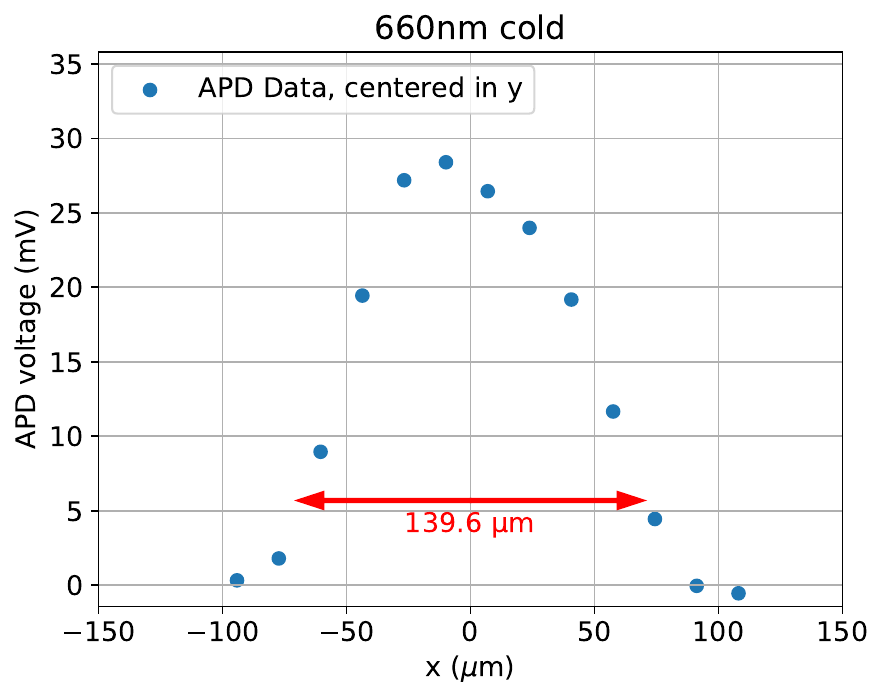}
    \caption{
    \textbf{Left}: Beam spot width along beam scan direction, as calculated from pulse-width measurements at 300~K with 660~nm photons. Distance units use the same calibration described in Figure~\ref{fig:APDspotSize660}. The 4$\sigma$ width is shown in red. \textbf{Right}: Corresponding results for the same system configuration when operated cold.}
    \label{fig:spotxwidth}
\end{figure*}
During warm testing, we noted a marked variability in observed pulse duration that was not strongly correlated with any one variable. Pulse width tests were performed with the system on the bench, installed in the cryostat (first open and then under vacuum), and even as a function of temperature during different stages of cooling to 10~mK. The quoted warm pulse widths correspond to the shortest stable values observed between 4~K and 300~K.

After cooling to 10~mK, the measured pulse widths decreased by 20--50\% for each wavelength, with the results for the lowest stable pulse widths at 300~K quoted in Table~\ref{tab:pulseWidth}. Rather than a gradual decrease, consistent with a continuous impedance change in the MEMS system, there was a discrete shift in performance when the system was heated from $<1$~K to 4~K. We attribute this change to the aluminum control lines on the MEMs mirror going through their superconducting transition, resulting in increased drive bandwidth, as the RC network of the MEMs mirror becomes purely capacitive.

We can use the detailed spot size measurements acquired with the 660~nm diode to back out the effective change in MEMs scan speed, removing the optical effects we have described earlier and compensating for any focal length drift. The measured pulse width decreased by $\sim$40\%, {\it i.e.} far more than the observed $\sim$3\% decrease in RMS spot size. For a more accurate comparison, we calculated the change in spot width along the scanning direction (X) without removing the effect of the output fiber. As seen in Figure~\ref{fig:spotxwidth}, the result for 660~nm was a larger total decrease of $\sim$10\%, still less than the observed decrease in pulse width. The remaining effect after compensating for the spot size change thus reflects the change in mirror speed.


By comparing the the 660~nm pulse width to spot size in the scanning direction, we estimate linear scan speeds of 24.5~m/s warm and 31.6~m/s cold. Including the scanning lever arm ($l_a=84$~mm), we convert these scan speeds to angular velocities of $\xi\sim$ 292~rad/s warm and 376~rad/s cold. Although full measurements were only done for 660~nm, the resulting scan speeds should be wavelength-independent.

If we further assume that the mirror sweeps through its full angular range, we can estimate the effective oscillation frequency the mirror attains during the drive signal and compare this to the resonant frequency quoted for the system when warm. From Appendix~\ref{app:timing}, we find that:
\begin{equation}
    v_{max} = l_a \xi = l_a 2 \theta_m \omega
\end{equation}
which implies for $2\theta_m = 10~\mathrm{degrees} = 0.175~\mathrm{rad}$ an oscillation frequency of $\omega/2\pi \sim$265~Hz warm and 341~Hz cold. This result essentially implies we are driving the mirror as fast as mechanically possible.

\section{Discussion}

This paper presents a detailed device-agnostic characterization of an enhanced cryogenic MEMS steering and optical chopping system over the range 365~nm to 970~nm. While the electrical set-up and enclosure designs are identical to those in the original system described in Ref~\cite{stifter2024}, we have demonstrated in the lab substantial improvements to the optical focuser design. As a result, some key system capabilities have improved by almost an order of magnitude in wavelength, and we have demonstrated spot sizes smaller than 100~$\mu$m across this entire spectral range. We also demonstrated robust optical chopping capability across the majority of the spectral range with performance limited by the mechanical properties of the MEMS mirror and its demonstrated optical behavior.

While not discussed in detail here thus far, one of the main advantages of this technique is the high degree of control fidelity achievable with practically zero power dissipation. For all of the measurements described in this paper, MEMS-related heating was typically negligible, and full-scale chopping modulation was retained as we added filtering to the MEMS control circuit. For these measurements, the observed heating was consistent with the expected heat loads from optical absorption in the system, including the fibers, and did not correlate with MEMS mirror speed or driving frequency. Forthcoming measurements using this second-generation MEMS scanning system with phonon-sensitive KIDs\cite{Temples_2024} and charge-sensitive superconducting devices should confirm the negligible thermal impact this versatile characterization technique has on real-world device performance.

Moving forward, we anticipate making specific further improvements to the system that will enable a broader range of cryogenic imaging capabilities. 
First, we have completed the design of a smaller system that does not require a secondary
flat mirror to achieve the same scan area and beam spot size of the present system. 
The new design uses optics with smaller focal lengths than used now, and is
compact enough so that the majority of the beam extends
outside the primary enclosure.  This system will achieve the same spot sizes with the optical components used in this system, and should be capable of diffraction-limited 
spot sizes $<$ 10 $\mu$m for the shortest scan range and a single-mode
fiber input. The beam modulation components will be housed in a package 
roughly 50~mm x 25~mm x 25~mm in volume that will couple to a single-mode output fiber.
This new system will be useful for producing few~$\mu$m-scale mapping of small-featured devices, such as semiconducting spin qubits, and will be mountable inside the bore of a standard cryogenic magnet.


Second, our long-term goal is to realize MEMS scanning across the full range of wavelengths supported by commercial reflective optics. The current system
is fundamentally limited not by the mirrors and the MEMS performance, but by the multimode fibers and their mode structures. Operation at $\sim$ single-$\mu$m wavelengths will involve using single-mode IR fiber optics. We plan to test such a configuration on the bench before it is deployed by collaborators in a current (v.2) scanner. For longer wavelengths, we plan to couple the focusing unit to mm-wave capillary waveguides, with the goal of achieving mm-scale spot sizes that can be used to characterize cm-scale arrays in conjunction with 4~K photon sources. This approach will enable higher-fidelity ground-based testing of mm-wave imaging arrays and {\it in situ} dark-box testing with a suitable photon absorber added to the MEMS system. It also will allow us to steer ambient photons coupled to the waveguide away from the device under test.

\begin{acknowledgements}
    The MEMS mirror used for this device was fabricated by Mirrorcle Technologies under an NRE contract with FNAL, SLAC, and NIST. Ansys Zemax OpticStudio 2025 R1 software was used for optical simulations. This work was supported in part by a KIPAC Innovation Grant, which supported initial undergraduate work on the reflective focuser by AN and GP. KS and NT were supported by a SLAC LDRD led by KS. TA was supported in part by the U.S. Government under ARO grant W911NF-22-1-0257, with work led by NK. NT was supported in part by the KIPAC postbaccalaureate fellowship program. RC was supported by a Geoff and Josie Fox Fellowship. BY was supported by a Lee and Seymour Graff Endowed Chair fund. This manuscript has been co-authored by FermiForward Discovery Group, LLC under Contract No. 89243024CSC000002 with the U.S. Department of Energy, Office of Science, Office of High Energy Physics. The work is supported by DOE, Office of Science, National Quantum Information Science Research Centers, Quantum Science Center and Q-NEXT. 

    The authors 
    thank Britton Plourde, Bevin Haung, and Yaniv Rosen for feedback on system use/requirements for work related to this paper that informed some of the testing described here.

    NK, BY, and BC dedicate the work done for this paper to Sae Woo Nam, who contributed substantially to quantum detector technology related to this project, and passed away tragically before we were able to work together to realize the final product. His innovative spirit is infused in this project, and we are grateful for the insight he shared when discussing the program to build these devices. He is sorely missed.
\end{acknowledgements}

\bibliography{report}   
\bibliographystyle{spiejour}   

\appendix

\section{Description of Experimental Setup}\label{app:expSetup}

The device tests described in the main text are the result of substantial trial and error with cryogenic fiber optics, which we endeavor to document in this appendix section. In particular, designing broad-band cryogenic fiber optics with low enough loss to perform loop-back measurements was more difficult than anticipated. 

\begin{table*}[hp]
    \centering
       \caption{Off the shelf components used for the tests described in this paper. All fibers are SMA connectorized with vacuum-compatible SMA connectors sourced by Thorlabs.}
    \begin{tabular}{|c|c|c|}
    \hline
        Component & Part Used & Specifications \\
        \hline
         Reflective Collimator & ThorLabs RC08 & 33~mm RFL, 0.15 NA maximum, 11mm aperature\\
         Off-Axis Parabolic Mirror & Thorlabs MPD189-P01 & 203.2 mm RFL\\
         Input Optical Fiber & ThorLabs FG010LDA & 10$\pm$3 $\mathrm{\mu m}$ core diameter, 0.10 NA\\
         Output Optical Fiber & Thorlabs FG105LVA & 105$\pm$3 $\mathrm{\mu m}$ core diameter, 0.10 NA\\
         Fiber Jacketing & FT900KK & Black fiber jacket \\
         Fiber Vacuum Feedthrough & Accuglass FO1UV-2-K40 & 105~$\mu$m-diameter UV/VIS KF40 feedthrough \\
         Avalanche Photodiode & Thorlabs APD410A2 & 25 A/W peak response, 12.4 V/$\mu$W \\
         \hline
    \end{tabular}
    \label{tab:ots_parts}
\end{table*}

\begin{table*}[hp]
    \centering
       \caption{Fiber transmission testing conducted to establish the optical path for the MEMS throughput and chopping tests described in this paper. 625~nm LED used unless specified otherwise. Measurements marked $^*$ indicate measurements made with the APD instead of the optical power meter. Switch from Accuglass to Thorlabs fibers made in August 2024 for fibers other the the 10~$\mu$m fiber, which was always the same Thorlabs sourced multimode fiber.}
    \begin{tabular}{|c|c|c|c|c|c|}
    \hline
        Test Date & Diode Current & Power & Temperature & Fiber Breaks & Experimental Conditions \\
        \hline
        2/15/24 & 550 mA & 128 $\mu$W & 300~K & 1 & Dunk test, control, UV/VIS Fiber \\
        &  & 126 $\mu$W & 300~K & 1 & Dunk test, feedthrough in DR plate, UV/VIS Fiber \\
        &  & 119 $\mu$W & 70~K & 1 & Dunk test, feedthrough in DR plate, UV/VIS Fiber \\
        \hline
        2/22/24 & 500 mA & 12.6 $\mu$W & 300~K & 13 & DR open. Loop-back at 10~mK stage. \\
        & 300 mA & 7.6 $\mu$W & 300~K & 13 & DR open. Loop-back at 10~mK stage. \\
        & 500 mA & 38.7 $\mu$W & 300~K & 6 & DR open. UV/VIS transmission to 10~mK \\
        & 500 mA & 49.2 $\mu$W & 300~K & 6 & DR open. IR/VIS transmission to 10~mK \\
        \hline
        2/23/24 & 500 mA & 12.6 $\mu$W & 300~K & 13 & Vacuum. Loop-back at 10~mK stage. \\
        2/26/24 & 500 mA & 9.18 $\mu$W & 37~mK & 13 & Vacuum. Loop-back at 10~mK stage. \\
        \hline
        6/12/24 & 100 mA & 30.3 $\mu$W & 300~K & 0 & DR open, IR/VIS Direct Power Measurement \\
        & 100 mA & 9.3 $\mu$W & 300~K & 3 & DR open, IR/VIS transmission to 4~K \\
        & 100 mA & 6.3 $\mu$W & 300~K & 5 & DR open, IR/VIS transmission to 10~mK \\
        6/13/24 & 1~A & 193~nW & 300~K & 6 & DR open, IR/VIS loop back, 10~$\mu$m fiber \\
        \hline
        6/24/24 & 900 mA & 160~nW & 300~K & 6 & DR closed, IR/VIS loop back, 10~$\mu$m fiber \\
        6/26/24 & 900 mA & 120~nW & 4~K & 6 & DR closed, IR/VIS loop back, 10~$\mu$m fiber \\
        \hline
        August 2024 & 850 mA & 1.14~mW & 300~K & 0 & 200~$\mu$m Input fiber \\
        & & 105~$\mu$W & & 0 & 105~$\mu$m Input fiber \\
        & & 66~$\mu$W & & 1 & 200~$\mu$m + 105~$\mu$m fiber  \\
        & & 43~$\mu$W & & 1 & 200~$\mu$m + 105~$\mu$m fiber + Feedthrough \\
        & & 622~nW & & 1 & 200~$\mu$m + 10~$\mu$m fiber + Feedthrough \\
        \hline
        10/23/2024 & 1~A & 2.4~nW$^*$ & 10~mK & 3 & DR Closed, MEMS Chopping Throughput \\
        11/11/2024 & & 97~nW$^*$ & 300~K & 3 &  \\
        \hline
        December 2024 & 850~mA & 92.5~$\mu$W &  300~K & 1 & 365~nm LED, 200~$\mu$m + 105~$\mu$m fiber  \\
        & & 91.7~$\mu$W & & 1 & 365~nm LED, 105~$\mu$m + 200~$\mu$m fiber \\
        \hline
        1/31/25 & 1~A & 29.5~$\mu$W & 300~K & 6 & 660~nm LED, IR/UV/VIS loop back + Thorlabs 105~$\mu$m fiber \\
        2/4/25 & 1~A & 1.23~$\mu$W & 4~K & 6 & 660~nm LED, IR/UV/VIS loop back + Thorlabs 105~$\mu$m fiber \\
        \hline
    \end{tabular}
    \label{tab:fiberLoss}
\end{table*}

\subsection{Optical Throughput Testing - Details}

The use of optical fiber for cryogenic measurements is not a novel concept (see e.g. Refs~\cite{Moffatt_2016,Romani_2018,Stanford_2021,Temples_2024}). However, previous experiments in our group had always focused on delivery of a highly attenuated photon beam to single-photon sensitive readout, or at most replicating keV-scale energies with optical energies. In the past we relied on having a fiber break at each thermal stage, plus an optical filter at the cold plate stage, to reduce the mW-scale energy at room temperature to pW-scale pulses at 10~mK. This filtering also served to block 300~K radiation from reaching the payload.

For this measurement, we needed to first demonstrate that we can transmit an optical signal from 300~K to 10~mK and back without losses generating substantial heating at the base stage of the dilution refrigerator (DR). Our typical fiber setups used vacuum compatible fibers (UV/VIS or IR/VIS depending on the application) from AccuGlass with a 105~$\mu$m core and a PEEK jacket. The fibers were installed with breaks at every stage, which contributed non-trivial losses. As a basis for comparison, we had Oxford install long, unbroken multi-mode fibers without explicit breaks from 300~K to the base stage, which were delivered with our cryostat. Our intent was to characterize the loss from having multiple breaks as a function of fridge temperature. 

We initially used Thorlabs fiber-coupled LEDs and a PM16-130 USB optical power meter to measure fiber throughput at a handful of optical wavelengths between 280 and 990~nm. These LEDs are generally rated for about 1~mW ouput into a 200~$\mu$m core multimode fiber, with variations depending on the LED wavelength. Direct coupling of a 200~$\mu$m fiber into the power meter yielded $\sim$1.7~mW for an equivalent driving current of 1~A in constant current mode. Using this result as the reference value for zero loss showed that we got a reasonably high coupling efficiency into the 200~$\mu$m fibers we use for coupling to 300~K, relative to the expected power levels for these diodes. 

Our initial warm round-trip measurements through the Oxford-installed fibers showed $\sim$7.2 $\mu$W/A round-trip transmission at 300~K, which quickly went to zero upon cooldown. In comparison, the typical Accuglass chain, which includes of order 11 SMA connections inside the cryostat, showed a round trip power level of 25~$\mu$W/A warm, dropping to 20~$\mu$W/A after the cryostat reached base temperature (see Table~\ref{tab:fiberLoss}). This amount of loss (roughly an attenuation of 20~dB) was expected due to fiber mismatch at each cryogenic connection, but we expected the single-fiber path to have much higher throughput overall. We eventually discovered that the FC/PC connectors used in the original single-fiber installation were highly unstable when cooled. Direct loop-back tests of the connectorized fiber using liquid nitrogen (LN) showed zero received power well before 77~K was reached.

We then attempted to characterize our losses. First, both fibers have core diameters of 100~$\mu$m, which can account for the first loss of at least 75\% (6~dB) in coupling efficiency, moving from the 200~$\mu$m to 100~$\mu$m cores. In fact, the loss was found to be greater than that because of numerical aperture and reflection effects. For example, our 300~K throughput tests showed $\sim$~100~$\mu$W/A compared to the expected 1.7~mW/A. This means the remaining connectors contribute about 12~dB of loss warm, and about 13~dB cold. We found that the majority of the reduction happened above 77~K, meaning liquid nitrogen (LN) dunk tests could be used to establish expected performance near 0~K.

To better model expected loss contributions, we then conducted a series of dunk tests with the SMA-connectorized Accuglass fibers. The best optical transmission with an air-side fiber coupler was found to be 232$\mu$W/A. The cryogenic fiber coupler in the feedthrough plate provided 225$\mu$W/A at 300~K, and 216$\mu$W/A in LN vapor, without any further drop in transmission upon immersion in LN. That is a relative loss of 0.3~dB between cold and warm measurements. We were also able to see roughly comparable shifts in loss simply by bending the warm cable. This suggests that the SMA interconnects are temperature stable and that all 11 interconnects combined are likely to contribute only $\sim$4~dB of loss. The UV/VIS fibers are rated for $\sim$ 0.05~dB/m of loss and thus contribute $\sim$ 0.2~dB to the total loss in our opitcal path. 

The fibers themselves have an expected interconnect loss specified by Accuglass of up to 2~dB/connection; this implies that our measured loss is dominated by fiber interconnect loss in the systems with many fiber breaks, and not by thermal or internal losses in the fibers themselves. Extensive bench testing was done to optimize overall throughput cold, given that in this test we expected to be limited by our ability to re-focus the beam back into the output fiber. As part of this testing, we switched to the lower NA (0.10) Thorlabs fiber listed in Table~\ref{tab:ots_parts} to get a smaller core diameter and stay within the limit of the reflective collimator specifications. We also switched the output fiber to be of the same material as the new input fiber. Our testing indicated that this smaller NA fiber had higher interconnect loss than the previous fiber, and we thus opted for long, single-fiber runs to and from the 10~mK cryostat stage to minimize overall system loss. Bench top tests of the new fiber are summarized in Table~\ref{tab:fiberLoss}, including a 4~K measurement demonstrating little change in transmission at cryogenic temperatures.


\begin{figure}[!th]
    \centering
    \includegraphics[width=0.87\linewidth]{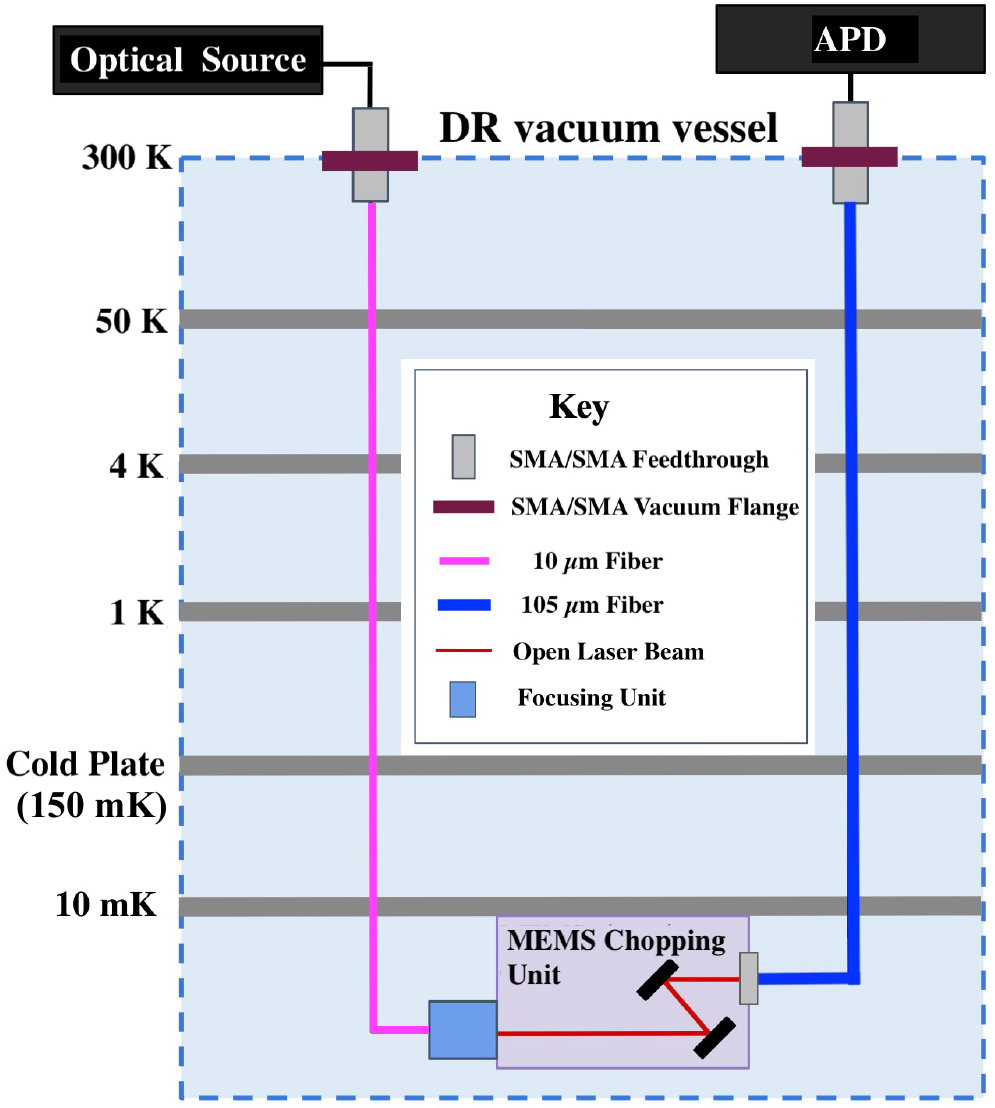}
    \caption{ 
    Chopping test setup in the dilution refrigerator. Light from an LED outside the vacuum vessel is sent through a 10~$\mu$m-diameter fiber (pink) to the base stage of the cryostat where the MEMS chopping unit is mounted. After chopping, the light beam (red) is collected and redirected to the top of the cryostat using a 105~$\mu$m-diameter fiber (blue). The optical output signal is read out with an APD at 300~K.}
    \label{fig:fridgeDiagram}
\end{figure}

The final configuration of fiber optics used for the pulse timing and cryogenic spot size measurements described in the main text is shown in Figure~\ref{fig:fridgeDiagram} with components listed in Table~\ref{tab:ots_parts}. Throughput for an equivalent system was measured on the bench in August 2024, with key results summarized in Table~\ref{tab:fiberLoss}. In the cryostat, we expected the input path to have a cold attenuation on the order of 35~dB relative to power injected at the 300~K vacuum feedthrough. We expected only a small amount (3-5 dB) of loss in the output path, given that the output fiber diameters increase from 10~$\mu$m to 105~$\mu$m to 200~$\mu$m in moving from 10~mK to 300~K.
These values are important for understanding the cryogenic trends seen in the scanning measurements discussed in the main body of the text.

\subsection{APD Data Acquisition and Throughput Measurement - Details}

As described in the main text, to measure optical pulse characteristics, we used a Thorlabs APD with 400~MHz bandwidth and maximum conversion gain of 12.4 V/$\mu$W at a peak wavelength of 600~nm. The setup, shown in Figure~\ref{fig:fridgeDiagram}, was optimized based on our experience commissioning the pulse measurement system on the bench. Rather than utilize a fiber-coupled APD, which has a bandwidth and performance defined by refractive optics, we opted for a free-space APD with an SMA fiber adapter to allow the fiber to shine freely on the APD. As a result, we don't expect the coupling to be highly efficient, but the APD output with our nominal fiber setup was found to drive the APD halfway to saturation as it was. We therefore did not perform any additional efficiency enhancements prior to the testing described here, and left the APD set to its maximum gain setting.


The results of this testing are summarized in the main text, with power throughput summarized for comparison to the power meter in Table~\ref{tab:fiberLoss}. Our best throughput in warm (300~K) testing of the MEMS setup, with the spot aligned at the maximum responsitivity point, was a power of about 100~nW. This is only about 20~nW lower than the loopback test using the 10$\mu$m fiber in a 4~K test, suggesting that additional losses in the warm APD setup are small, and coupling efficiency into the output fiber is fairly high. An excess loss of 16~dB was measured compared to an expected loss of 2~dB or less, which we now attribute to substantially higher than expected loss in the 105~$\mu$m Thorlabs output fiber.

\section{Expected Optical Performance}\label{app:timing}

This paper describes the cryogenic performance of a new optical scanning and chopping unit that uses a broad-band optical focusing apparatus in place of the refractive focuser used previously Ref~\cite{stifter2024}. The new design, with its separate focusing and collimation optics, offers flexibility to define the scan range, spot size, and chopping-pulse timing during scanner construction. Below, we show in detail how specific component choices affect these three MEMS-scanner metrics.

\subsection{Spot Size}

First, consider the case where a perfectly collimated monochromatic source is focused to a diffraction-limited spot at its focal length, a distance $f_0$ from the focusing optic. After traveling a distance $l$, the beam is incident on a movable MEMS mirror, which can cover its full angular deflection in some time $t$, sweeping the beam across a focal plane at distance $f_0-l$ from the MEMS mirror. This setup is pictured in Figure~\ref{fig:diagram} in the main text.

\begin{figure*}
    \centering
    \includegraphics[width=0.5\linewidth]{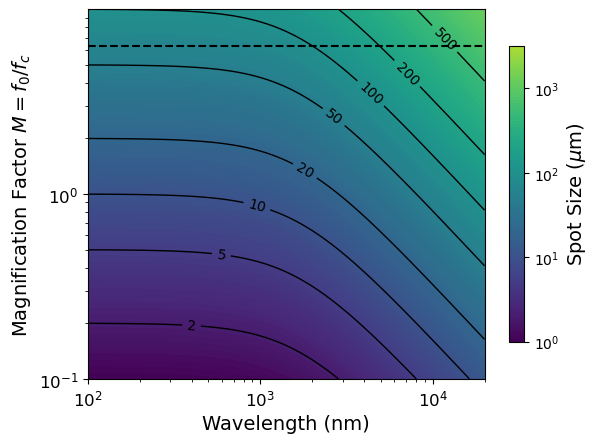}
    \includegraphics[width=0.48\linewidth]{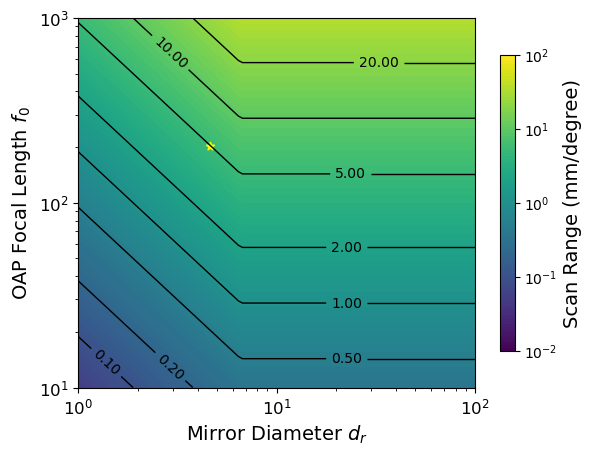}
    \includegraphics[width=0.5\linewidth]{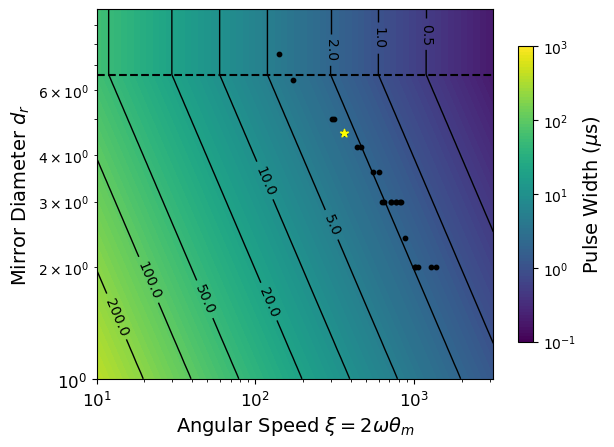}
    \includegraphics[width=0.48\linewidth]{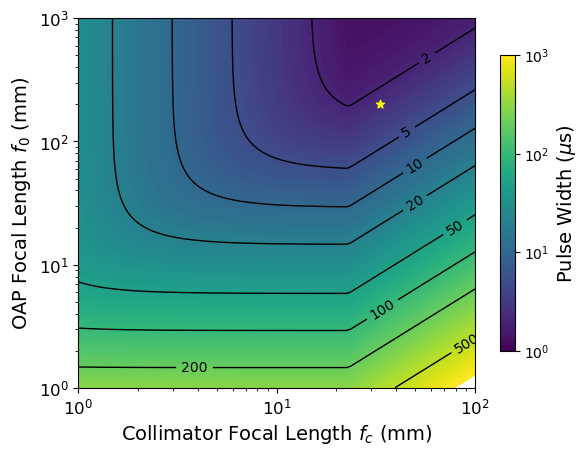}
    \caption{Design expectations for the scanning and pulse generation performance of this device for a range of tunable parameters. \textbf{Top Left}: Spot size at the focus versus wavelength and magnification factor $M=\frac{f_0}{f_c}$. The horizontal line is the magnification factor for the design discussed in this paper. \textbf{Top Right}: Scan range per degree of MEMS mirror deflection as a function of OAP focal length and MEMS mirror size, with parameters for this device indicated with a star. \textbf{Bottom Left}: Pulse width for fixed optics as a function of MEMS mirror parameters, with mirror diameter on the y-axis and the frequency-angle product of the mirror on the x-axis. The mirror used for these results is shown with a star. The range of mirrors available from Mirrorcle are shown for reference as black dots. \textbf{Bottom Right}: Pulse width for fixed MEMS mirror, as a function of collimator and main OAP focal lengths, assuming 10~$\mu$m input fiber and 105~$\mu$m output fiber.}
    \label{fig:spotsize_design}
\end{figure*}

Consider a simple model for the beam waist size when diffraction-limited at the focal spot. The diameter $d_o$ of the beam, assuming perfect collimation, is given by:
\begin{equation}
d_0 =2 f_{c} \cdot NA
\end{equation}
where $NA \sim n \sin \theta$ is the fiber's numerical aperture, $n$ index of refraction, and $f_{c}$ is the focal length of the reflective collimator mirror. For a diffraction limited spot, the size of the central diffraction peak can be defined as:
\begin{equation}
d_{diff}=(1.22)\frac{\lambda f_{0}}{d_0}=\frac{f_{0}}{f_{c}}\frac{1.22\lambda}{2NA}
\end{equation}

In practice, the divergence of the collimated beam limits longer focal length optics, and we need to include the impact of beam divergence on the spot size. For the fiber-based reflective collimators we use, the divergence angle of the beam is given as:
\begin{equation}
    \theta_{c}=\frac{d_{f}}{f_{c}}
\end{equation}
where $d_f$ is the core diameter of the input optical fiber. To calculate the spot diameter at the focal length of the focuser from a beam with a known divergence angle,
\begin{equation}
    d_{div}=\theta_c f_0 = d_f \frac{f_0}{f_c}
\end{equation}
For gaussian beams, we expect the divergence and diffractive effects to add in quadrature, so we have the effective spot size at the focus of
\begin{align}
  d_{eff} &= \frac {f_{0}}{f_{c}}\sqrt{d_{f}^2+(1.22\lambda/2NA)^2} \label{eq:spotSize}\\
  &= \frac{f_{0}}{d_0}\sqrt{(2 NA \cdot d_{f})^2+(1.22\lambda)^2} .
\end{align} 
In the limit that the fiber is much larger than the diffraction limited spot, this is just a magnifier, where an image of the fiber is projected at the focal point of the focusing optic, scaled by the magnification ratio $M=f_0/f_c$. 

\subsection{Scan Range}

For scanning applications, we care about both the spot size and scan range. The beam waist varies with distance from the main focusing mirror as $d = d_0\frac{f_0-l}{f_0}$ where $d_0$ is the initial size of the collimated beam. For a mirror with reflector of size $d_r$, the maximum distance from the focal spot is:
\begin{equation}
    l_a = (f_0-l)_{max} = \frac{d_r}{d_0} f_0 = \frac{d_r}{2 NA}\frac{f_0}{f_c}
\end{equation}
All mirrors considered here have small angular deflections, and we can thus approximate the scan range distance as:
\begin{equation} \label{eq:scanRange_app0}
    d_{scan} \approx 2\theta_d l_a = \frac{d_r \theta_d }{NA}\frac{f_0}{f_c}
\end{equation}
where $\theta_d$ is the angle the beam is reflected. The beam angle is doubled for a given angular displacement of the mirror $\theta_m$, so we find $\theta_d = 2\theta_m$, and:
\begin{equation} \label{eq:scanRange_app}
    d_{scan} \approx \frac{2 d_r \theta_m }{NA}\frac{f_0}{f_c}
\end{equation}
Here again, we find the magnification factor M = $\frac{f_0}{f_c}$, meaning that scan range and spot size scale proportionally, for all other system parameters fixed.

For a MEMS mirror with diameter equal to or larger than the initial collimated beam size, the scan range is just:
\begin{equation}
    d_{scan} \approx 2\theta_d f_0 = 4\theta_m f_0,
\end{equation}
which implies that for a setup not limited by the mirror diameter, we can reduce spot size by increasing the focal length of the collimating optic. 

\subsection{Pulse Duration}

Next, we use the above to compute the effective pulse duration for a mirror swept across a slit much smaller than the beam spot. For maximum deflection angle $\theta_d$, the mean scan velocity scales as 
\begin{equation}
    v_{beam} = \frac{d_{beam}}{t} \approx \frac{2(f_0-l)\theta_d}{t}
\end{equation}
where $d_{beam}$ is the diameter of the beam at the output focal length and $t$ is the average transit time for the mirror to pass through the scan range. 

The scan velocity at the center of the beam path depends on the spring constants of the MEMS mirror, mirror mass, and the driving voltage, which we can approximate given known mirror properties, but is unknown for the experimental conditions we intend to run the device at. When driven harmonically with a frequency $\omega_0$, the scan position is given as 
\begin{equation}
    x(t) = l_a\theta_d sin(\omega_0 t),
\end{equation}
and we expect the velocity to be highest when the mirror is centered at $x=0$, giving a maximimum scan velocity of 
\begin{equation}
    v_{max} = \left.\frac{d}{dt}x\right|_{t=0} = l_a\theta_d\omega_0 = l_a \xi
\end{equation}
We thus take $\omega_0$ as an effective angular speed for the mirror, which in principle is correlated with deflection angle for larger drive amplitude. We will see later that the product $\xi=\theta_d\omega_0 = 2\theta_m \omega_0$ is inseparable in our pulse timing calculation, and should be the relevant parameter to compare mirrors.

Using the results from the previous section, we thus get the scan speed
\begin{equation}
    v_{max} = f_0 \xi \frac{d_r}{d_0} = \xi \frac{d_r}{2NA} \frac{f_0}{f_c}
\end{equation}
Thus scan speed is determined by intrinsic mirror characteristics (frequency and mirror size, which in general are anti-correlated) as well as the area and focal length of the focusing optics.

The expected pulse duration is given by $t_{pulse} = d_{pulse}/v_{max}$. Using equations derived earlier, this gives a diffraction limited pulse width of
\begin{equation}
    t_{pulse,diff} = \frac{1}{f_{c}\xi}\frac{1.22\lambda}{2 NA}\left(\frac{d_0}{d_r}\right) = \frac{1}{\xi}\frac{1.22\lambda}{d_r}
\end{equation}
This is the smallest pulse that a mirror with given $\xi$ and $d_r$ can produce for a very small sensor, assuming mirror size is smaller than the initial beam diameter.

Including divergence effects as in the previous section, we find an effective pulse duration for a realistic geometry of
\begin{align}\label{eq:fullTiming}
    t_{eff} \approx \frac{d_{eff}}{v_{max}} = \frac{\sqrt{(2NA\cdot d_{f})^2+(1.22\lambda)^2}}{\xi d_r}, \; d_r \leq d_0
\end{align}
The pre-factors here are now the same as the previous calculation, but the fiber diameter will dominate the timing if it is substantially larger than the wavelength of the light. Single-mode fibers will therefore always produce diffraction-limited chopping in the optical regime, but we are focused on producing a broad-band design, which requires light guides to admit multiple modes; this restricts some of our optimization to utilize larger collimators and longer focal lengths. 

We note that the scaling with mirror size only holds until $d_{r} \ge d_0$, after which $d_0$ becomes the limiting factor. This leads to some confusing performance in the case of small numerical aperture, as the above derivation assumes this also allows the mirror to be placed further from the slit, canceling out the increased spot size due to diffraction effects. In this case, our maximum velocity becomes
\begin{equation}
    v_{max} = f_0 \xi
\end{equation}
which gives the equation
\begin{equation}\label{eq:fullTiming_alt}
    t_{min} \approx \frac{\sqrt{(d_{f})^2+(1.22\lambda/2NA)^2}}{\xi f_c}, \; d_r \geq d_0
\end{equation}
This is the relevant behavior for a scan-optimized setup, where the MEMS mirror is located as close to the focusing optic as possible. Readily available MEMS mirrors are typically much smaller than focusing optics, and larger mirrors tend to have lower values of $\xi$, meaning that a realistic optimization needs to take into account the range of mirror sizes and $\xi$ values to determine which will best match timing requirements.

\begin{figure}
    \centering
    \includegraphics[width=\linewidth]{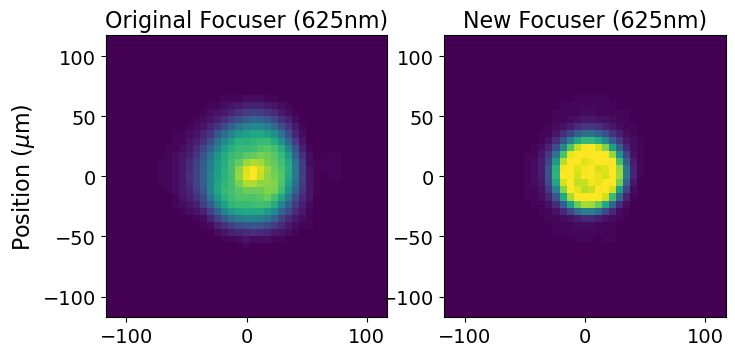}
    \includegraphics[width=\linewidth]{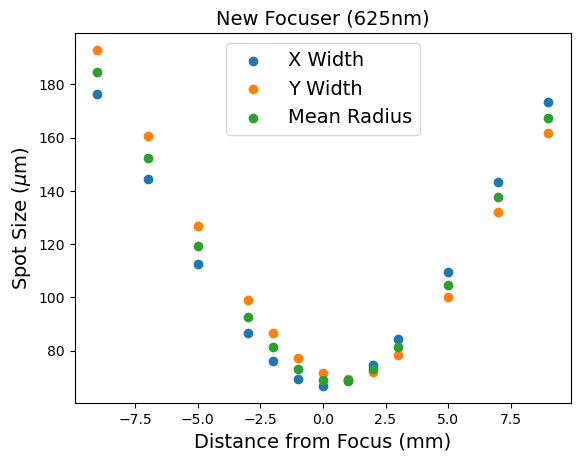}
\includegraphics[width=\linewidth]{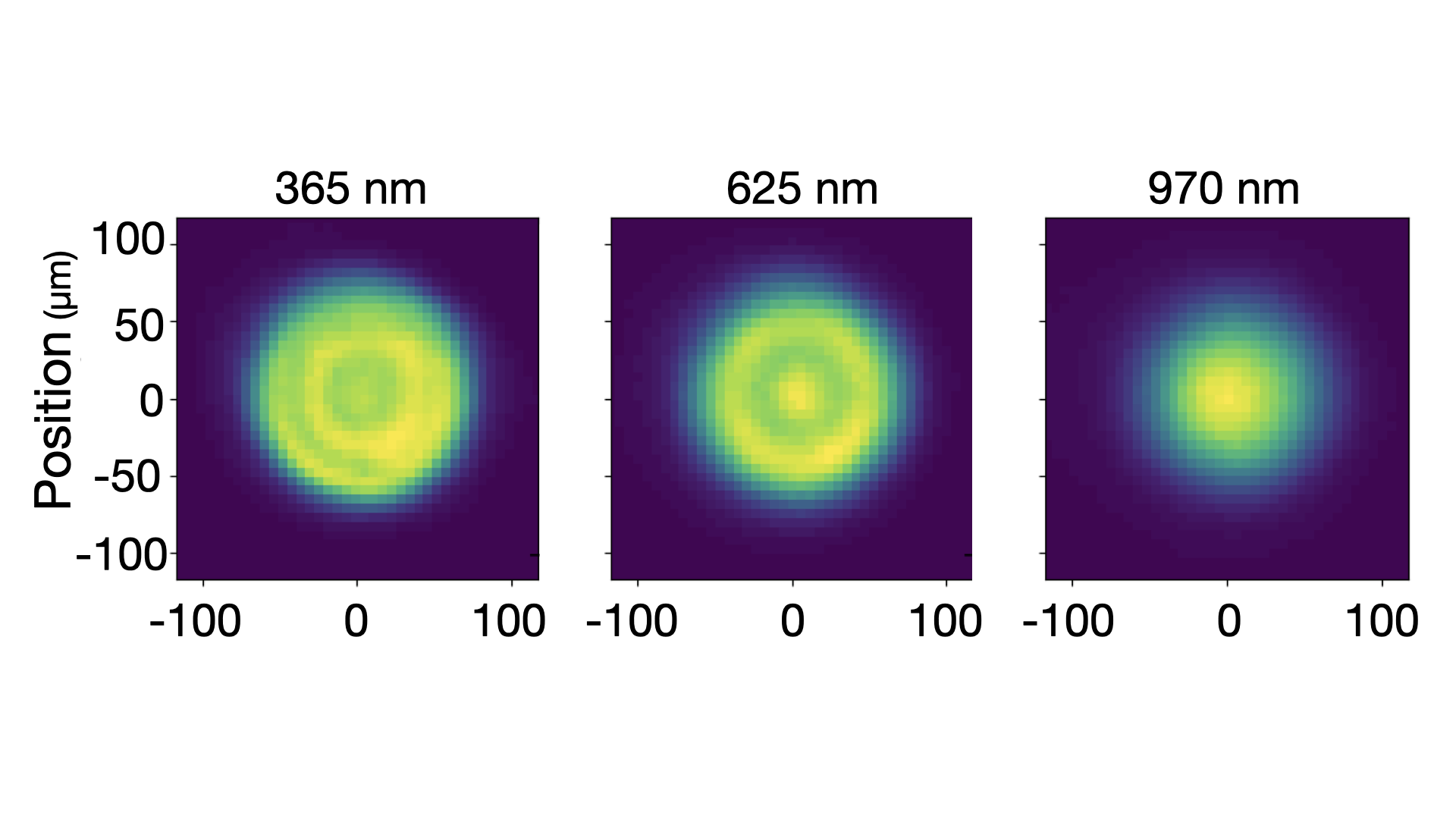}
    \caption{{\bf Top}: Comparison of CCD measured spot size for the copper multi-piece focuser assembly (left) and unibody 3d printed assembly (right), showing the smaller and more spatially resolved spot obtained with the new design. {\bf Middle}: Spot size vs z position in the beam, showing substantially reduced astigmatism and smaller spot size at the best focus, located at 1mm positive displacement in this measurement. {\bf Bottom}: Spot image with smaller collimator, producing a magnified spot on the CCD, as a function of wavelength. As wavelength increases, mode volume shrinks, and we can visibly see the reduction in number of modes as the spot goes from being dominated by a TEM20 mode at 365~nm to the TEM00 at 970~nm. The spatial resolution at the CCD within the spot, which is entirely limited by the diffraction-limited performance of the focuser, is also visibly better at short wavelengths, as expected.}
    \label{fig:newfocus}
\end{figure}

\section{System Performance Improvements}\label{app:newFocuser}

To date, we have produced and tested two versions of the fully enclosed MEMS scanner system, incorporating improvements based on initial cryogenic tests. The v1 design, described in detail in Ref~\cite{stifter2024}, achieved spot sizes of around 170~$\mu$m at fixed wavelength using a commercial refractive focusing system. The v2 design, for which the focuser described in this paper was designed, substantially improved the performance of the system, reducing spot size by a factor of 2 and introducing broad-band operation, allowing for warm interchange of light sources across a wavelength range of 300~nm to 1~$\mu$m limited only by optical fiber transmission cryogenically. The design shown in this paper was an intermediate design, using the focuser of the v2 system with modified v1 body, to demonstrate cryogenic performance. The remainder of the v2 system also decreased total copper volume to fit into a 5" diameter cryoperm can in order to test state of the art devices, like superconducting qubits, in an optimal environment. Results from this system are forthcoming with active cryogenic devices.

We are in the process of finalizing designs for a third version which will simplify alignment, reduce manufacturing cost, and improve reproducibility between devices. This will reduce the number of parts in the system, simplifying alignment, and allow for easier customization of magnification factor and scan range. The main improvement to the optics chain involves fixing the astigmatism in the focusing unit described in the body of the paper by manufacturing a unibody focuser, which allows for more precise angular alignment of the collimator and primary focusing mirror. The improvement in focuser performance is summarized in Figure~\ref{fig:newfocus}, where the astigmatism is drastically reduced compared to the existing system (see Figure~\ref{fig:spotSizeMeasured}), and the resulting spot size is reduced from 95 to 70 $\mu$m, with a 5 $\mu$m uncertainty.

As part of this investigation, spot measurements were carried out with a shorter focal length collimator (Thorlabs RC04), which magnifies the fiber by an additional factor of 2.2 for a total system magnification of 13.5. The resulting spot image is show in Figure~\ref{fig:spotSizeMeasured}, which more clearly shows the multi-moded nature of the light being emitted by the fiber. While the astigmatism-corrected spot sizes are within 20\% of the divergence-limited spot expectation, assuming an ideal flat emission profile for the fiber with a 10 micron width, in both the original data and the collimator with resolved astigmatism, we find the spot size gets smaller with increasing wavelength above roughly 700 $\mu$m. 

To explore this further, we can look at the magnified spot sizes for different wavelengths, shown at the bottom of Figure~\ref{fig:newfocus}. The most striking feature of these images is that the fiber mode is visibly changing as wavelength decreases. The number of modes in an optical fiber scales with wavelength as:
\begin{equation}
    N_{mode} \sim \frac{2\pi^2 a^2}{\lambda^2}NA^2,
\end{equation}
where $a$ is the radius of the fiber core.
For the wavelengths tested in this paper, the number of modes in the 10~$\mu$m-diameter optical fiber ranges from 150 at 365~nm to 20 at 970, resulting in a trend towards a more Gaussian profile as the fiber gets closer to behaving more like a single-mode fiber (defined as $N_{mode}\lesssim 3.1$). Detailed modeling of spot size vs. wavelength would thus require detailed mode modeling of the optical fiber, which is beyond the scope of this paper. We note, however, that the spot size is thus limited by fiber performance, and that smaller spots for a given scanning range are attainable by moving to single-mode fibers, at which point the diffraction and fiber limitations will need to be modeled in more detail.

\section{{\it In Situ} Spot Calibration}\label{app:spotCalibration}

As discussed in Section~\ref{sec:beamSize}, spot size was measured {\it in situ} by scanning across the output-plate fiber and recording APD voltage. The resulting data consisted of voltage as a function of MEMS setting. To convert the unitless MEMS setting to distance units, we independently fit the profile along perpendicular axes to a flat top (representing the output fiber) convolved with a Gaussian (representing the spot intensity), then scaled the flat-top width along each axis to match the known 105~$\mu$m fiber diameter. Instead of scaling along the MEMS x and y axes, we scaled along the first and second principal components, which indicate the directions in which the initial spot appears most and least stretched. An example of the conversion is shown in Figure~\ref{fig:transform}.

\begin{figure*}[!th]
    \centering
    \includegraphics[width=0.9\linewidth]{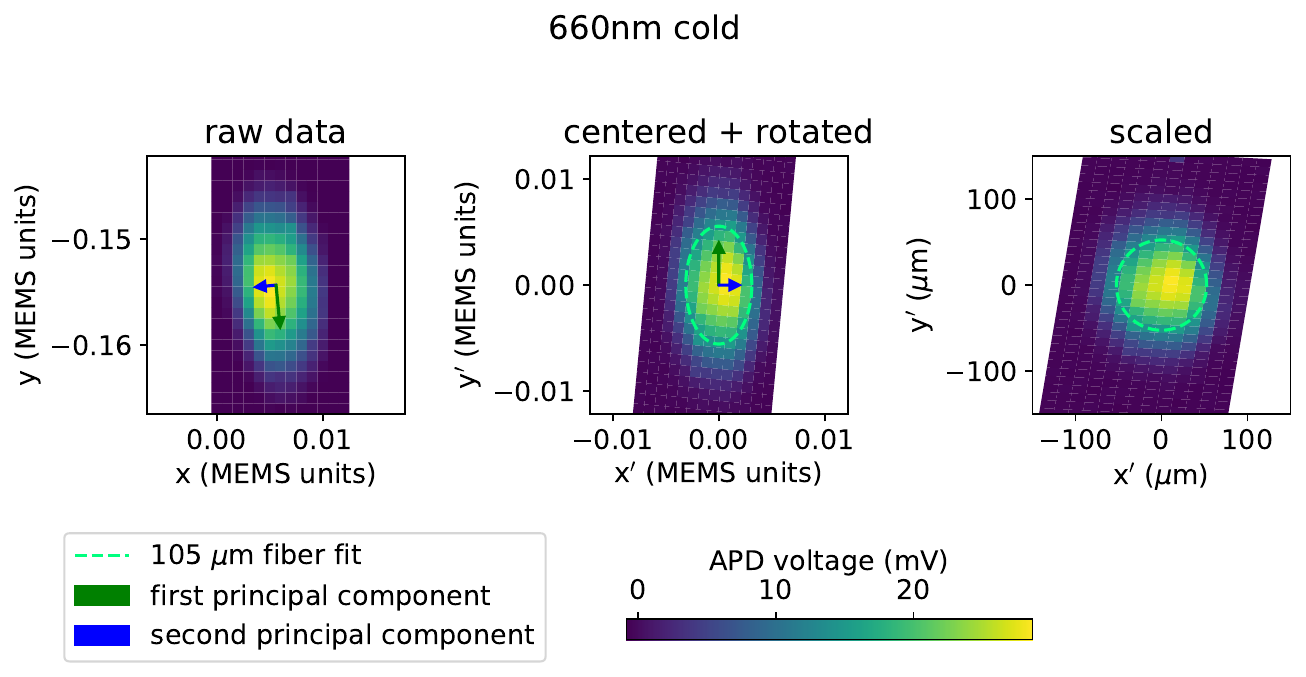}
    \caption{
    \textbf{Left}: APD output voltage (color bar scale) as a function of MEMS tilt-control settings for the cold 660~nm spot measurement shown in Figure~\ref{fig:APDspotSize660}. The first two principal components are plotted as vectors with length proportional to standard deviation along the relevant component (x or y). \textbf{Center}: The same data centered on the origin and rotated to use axes along the principal components. The fiber diameter in MEMS units is estimated by independently fitting the profile along each axis to a Gaussian convolved with a flat top and extracting the flat-top width. \textbf{Right}: The same data with each axis scaled so the flat-top width matches the fiber diameter of 105~$\mu$m.}
    \label{fig:transform}
\end{figure*}

\end{document}